\documentclass[a4paper,11pt]{article}

\usepackage{jcappub} % for details on the use of the package, please see the JINST-author-manual
\usepackage{lineno}
\usepackage{graphicx} % Required for inserting images
\usepackage{xspace}
\usepackage{hyperref}
\usepackage[normalem]{ulem}
\usepackage{float}
\usepackage{placeins}

\graphicspath{{./}{figures/}{figures/o4/}{figures/3g/}}

\usepackage[usenames,dvipsnames]{xcolor}
\newcommand{\swap}[1]{#1\xspace}

\def\code#1{\texttt{#1}}
\newcommand{\msun}{\ensuremath{M_{\odot}}\xspace}
\newcommand{\chieff}{\ensuremath{\chi_{\rm eff}}\xspace}
\newcommand{\chip}{\ensuremath{\chi_p}\xspace}

\newcommand{\thetajn}{\ensuremath{\theta_{JN}}\xspace}
\newcommand{\thetajl}{\ensuremath{\theta_{JL}}\xspace}
\newcommand{\vecL}{\ensuremath{\vec{L}}\xspace}
\newcommand{\vecJ}{\ensuremath{\vec{J}}\xspace}
\newcommand{\ans}{\ensuremath{a_{\rm NS}}\xspace}
\newcommand{\maxspin}{\ensuremath{0.4}\xspace}

\newcommand{\nsignals}{\ensuremath{60}\xspace}
\newcommand{\xgsmallestsnr}{\ensuremath{236.6}\xspace}

\title{Too small to fail: characterizing sub-solar mass black hole mergers with gravitational waves}
\arxivnumber{2305.19907}

\author[a,b]{Noah E. Wolfe,}
\author[a,b]{Salvatore Vitale,}
\author[a,b]{and Colm Talbot}
\affiliation[a]{LIGO Laboratory, Massachusetts Institute of Technology, 185 Albany St, Cambridge, MA 02139, USA}
\affiliation[b]{Department of Physics and Kavli Institute for Astrophysics and Space Research, Massachusetts Institute of Technology, 77 Massachusetts Ave, Cambridge, MA 02139, USA}

\emailAdd{noah.wolfe@ligo.org}
\emailAdd{salvo@mit.edu}

\abstract{
    The detection of a sub-solar mass black hole could yield dramatic new insights into the nature of dark matter and early-Universe physics, as such objects lack a traditional astrophysical formation mechanism.
    Gravitational waves allow for the direct measurement of compact object masses during binary mergers, and we expect the gravitational-wave signal from a low-mass coalescence to remain within the LIGO frequency band for thousands of seconds.
    However, it is unclear whether one can confidently measure the properties of a sub-solar mass compact object and distinguish between a sub-solar mass black hole or other exotic objects.
    To this end, we perform Bayesian parameter estimation on simulated gravitational-wave signals from sub-solar mass black hole mergers to explore the measurability of their source properties.
    We find that the LIGO/Virgo detectors during the O4 observing run would be able to confidently identify sub-solar component masses at the threshold of detectability; these events would also be well-localized on the sky and may reveal some information on their binary spin geometry.
    Further, next-generation detectors such as Cosmic Explorer and the Einstein Telescope will allow for precision measurement of the properties of sub-solar mass mergers and tighter constraints on their compact-object nature.
}

\begin{document}
\maketitle
\flushbottom % from JCAP template

\section{Introduction}

If some fraction of dark matter is composed of black holes, or gravitationally collapses to form black holes, gravitational waves (GWs) may offer the opportunity to directly probe the nature of dark matter.
Recent work has proposed that previous GW signals consistent with stellar-mass black holes can be sourced from black holes with a primordial origin \citep{Bird:2016dcv, Sasaki:2016jop, Clesse:2016vqa, Clesse:2020ghq}, however, there is currently no preference for those formation models over astrophysical channels \citep{Sasaki:2016jop, Clesse:2016vqa, Ali-Haimoud:2017rtz, Clesse:2017bsw, Wong:2020yig, DeLuca:2020qqa, DeLuca:2021wjr, Chen:2021nxo, Franciolini:2022tfm}.
For current- and next-generation ground-based gravitational-wave detectors like Advanced LIGO \citep{advancedligo}, Advanced Virgo \citep{advancedvirgo}, KAGRA \citep{kagra}, Cosmic Explorer \citep{cehorizonstudy}, and the Einstein Telescope \citep{einsteintelescope}, cleaner targets for dark matter searches may be compact object mergers involving a black hole with mass $\lesssim 1~\msun$.
The signals emitted by these events lie firmly in the frequency range accessible to ground-based detectors \textit{and} are immediately distinguished from traditional astrophysical formation channels by their mass alone\footnote{As their mass lies below the Chandrasekhar limit of ${\sim}1.4~\msun$ \citep{Chandrasekhar:1931ih}.}.
While microlensing surveys have placed constraints on the fraction of dark matter composed by $\mathcal{O}(1~\msun)$ black holes, the constraints may depend on the assumed mass distribution of the objects as well as the distribution of mass in the halo of the Milky Way \citep{Villanueva-Domingo:2021spv}.
Thus, it remains possible for some dark matter to be found in sub-solar mass black holes.
There are two categories of hypothesized sub-solar mass black holes: primordial black holes and dark matter black holes.

Primordial black holes (PBHs) could have formed from the gravitational collapse of over-densities in the early Universe \citep{zeldovich1967, hawking1971}.
Such objects have been proposed as a population of cold, collisionless dark matter \citep{Carr:2021bzv}.
The existence and mass distribution of primordial black holes depends strongly on their formation mechanism and underlying density power spectrum.
Based on horizon scale considerations, PBHs with a sharply peaked mass distribution near $\mathcal{O}(1~\msun)$ could have formed during the radiation-dominated era near the quark-hadron phase transition; the abundance of PBH masses may depend on the background cosmology \citep{Niemeyer:1997mt, Niemeyer:1999ak, Musco:2004ak}, equation of state of the early Universe \citep{Carr:2019kxo}, and the (non-)Gaussianity of the density fluctuations \citep{Young:2013oia, Bugaev:2013vba, Franciolini:2018vbk} (for a review of PBH formation in the radiation-dominated era, see \citep{Villanueva-Domingo:2021spv}).
PBH formation could have been driven by other physics during or beyond the radiation-dominated era (for a review of such scenarios, see \citep{Carr:2021bzv}).
Fluctuations generated by inflation could seed primordial black holes \citep{Dolgov:1992pu, Carr:1993aq, Ivanov:1994pa, Randall:1995dj, Garcia-Bellido:1996mdl}, which would be directly sensitive to the dynamics of inflation.
Other structures, like cosmic loops \citep{Hawking:1987bn}, bubbles of broken symmetry \citep{Khlopov:1999ys}, and domain walls \citep{Garriga:2015fdk} could also collide or collapse to form primordial black holes.
Additionally, the spectrum of PBH masses would be sensitive to accretion physics \citep{Villanueva-Domingo:2021spv}, and their formation into binaries depends on their clustering dynamics \citep{Raidal:2018bbj, Matsubara:2019qzv, Suyama:2019cst, Trashorras:2020mwn, Hutsi:2020sol, Phukon:2021cus} and their merger rate which are theoretically uncertain \citep{Clesse:2020ghq, Jedamzik:2020omx}.

Alternatively, black holes could form directly from the collapse of particle dark matter in certain dissipative dark matter models \citep{DAmico:2017lqj, shandera2018, Chang:2018bgx, Choquette:2018lvq, Latif:2018kqv, Essig:2018pzq,  Ryan:2021dis, Hippert:2021fch, Gurian:2022nbx}.
The precise mass distribution of such ``dark matter black holes" (DBHs) is strongly dependent on the microphysical details of the dark matter particles such as the possible dark matter species, their masses, and interaction cross sections.
The merger rate of DBHs would also depend on the larger-scale dynamics of dark matter halos.
Both the microphysical details and galactic-scale dynamics of dark matter are open questions \citep{deMartino:2020gfi}.
Ref.~\citep{shandera2018} studied a set of atomic dark matter scenarios that lower the Chandrasekhar mass with a heavier dark-proton analog, forming black holes $\lesssim 1~\msun$.
There, the dark matter microphysics is encoded in the dark matter cooling rates, and halo dynamics are encoded in the fraction of dark matter available to dynamically cool.

Thus, the discovery of a sub-solar mass black hole would probe the nature of dark matter and provide critical constraints on a rich theoretical landscape of physics in the early- and modern-Universe.
If such objects can form binaries and merge in the Hubble time, gravitational waves are the most promising method of detection.
Even the non-detection of sub-solar mass compact objects is already providing unique constraints on the parameter space of dark matter \citep{LIGOScientific:2022hai, Nitz:2022ltl, LIGOScientific:2021job, Nitz:2021vqh}.
However, fully realizing this promise hinges on positively identifying a compact object involved in a merger as (1) $\lesssim 1~\msun$ in mass and (2) a black hole.
This first question has previously been studied for astrophysical, super-solar mass black holes in next-generation detectors \citep[for example]{Pieroni:2022bbh} as well as sub-solar mass neutron stars \citep{Bandopadhyay:2022tbi}.
Given a gravitational wave signal from the coalescence of compact objects, the identification of a sub-solar mass black hole could be complicated by ``mimickers" such as sub-solar mass neutron stars or boson stars \citep{Liebling:2012fv}.
While such objects themselves would be astrophysically exotic \citep{2022NatAs...6.1444D, Landry:2021hvl, Meskhi:2021pdp} and potentially sourced from dark matter \citep{Giudice:2016zpa}, it is not yet clear how well current methods of gravitational-wave data analysis will distinguish between sub-solar mass black holes and these alternatives.
For example, when analyzing a low-significance sub-solar mass trigger, Ref.~\citep{Morras:2023jvb} could not exclude a neutron star origin for the sub-solar mass object.\footnote{See Ref.~\cite{Hannam:2013uu} for an analysis focusing on distinguishing light super-solar mass black holes from neutron stars.}

In this work, we estimate parameters for a set of simulated signals from sub-solar mass black hole mergers in current- and next-generation gravitational wave detectors to understand the feasibility of identifying sub-solar mass black holes with gravitational-wave signals across a range of binary black hole parameters.
First, we inspect the constraints we can achieve on the component black hole masses, to determine if we can confidently identify that a compact object is sub-solar mass in nature at all.
Then, we inspect two additional parameter sets for these signals-- the spins of the compact objects and the sky location of the binary-- both of which may rule out neutron stars in such signals.
We conclude with a discussion of our main findings.

\section{Methods}
\subsection{Gravitational-Wave Parameter Estimation}
To measure the source properties of gravitational-wave signals, we perform Bayesian parameter estimation.
Bayes' Theorem states that
\begin{equation}
    p(\theta | d) = \frac{\mathcal{L}(d | \theta) \pi(\theta)}{\mathcal{Z}(d)}
\end{equation}
where $p(\theta | d)$ is the posterior probability that the data $d$ contains a signal described by source parameters $\theta$, $\mathcal{L}$ is the likelihood of observing the signal given some source parameters, $\pi(\theta)$ is the prior probability of $\theta$, and $\mathcal{Z}$ is a normalization commonly referred to as the evidence.
In the case of gravitational-wave parameter estimation for emission from a quasi-circular black hole binary, $\theta$ includes 8 intrinsic parameters of the component black holes (their masses and spin vectors) and 7 extrinsic parameters of the binary (including its luminosity distance and location on the sky), for 15 parameters total.
We use the Whittle likelihood approximation in the frequency domain for the residual of the data minus the astrophysical signal\footnote{In other words, we treat the noise as stationary and Gaussian-distributed. Note this may not necessarily be true in next-generation detectors which will have many overlapping signals, but in principle, our results hold as e.g. one could subtract out this astrophysical ``noise" \citep{Wu:2022pyg}.}.
Here, the goal of our analysis will be to compute the posterior probability of $\theta$ for a series of simulated signals.
In this work, we use the nested sampling algorithm \citep{2004AIPC..735..395S, Skilling:2006gxv} implemented by \code{dynesty} \citep{Speagle:2019ivv} to estimate $p(\theta | d)$.

As the nested sampling algorithm iterates, we must evaluate $\mathcal{L}(d|\theta)$ which requires evaluating the waveform model at some proposal $\theta$. Signals from merging binaries with relatively low total mass or mass ratio will be relatively long; for the component masses listed in Table~\ref{tab:masses}, signals will be $\mathcal{O}(10^3)$ seconds in length compared to $\mathcal{O}(10^2)$ seconds for the larger total mass, near equal-mass binary neutron star GW170817 \citep{LIGOScientific:2018hze}.
This dramatically increases the number of frequencies at which we need to evaluate proposal waveforms for a given $\theta$, in evaluating $\mathcal{L}$, which is computationally expensive.
Instead, we use a ``heterodyned" likelihood \citep{Cornish:2021lje}, an approximation also known as ``relative binning" \citep{Zackay:2018qdy, Leslie:2021ssu}, which well-approximates $\mathcal{L}$ at far fewer frequencies by expanding it around its value at some fiducial parameters.
For simulated signals, we choose these parameters to be the true source parameters.
We also forego effects due to the rotation of the Earth, which would cause the antenna pattern to vary in time and increase the computational expense of $\mathcal{L}$.
We carry out parameter estimation using \code{bilby} \citep{Ashton:2018jfp, Romero-Shaw:2020owr} which implements a heterodyned likelihood as in Ref.~\citep{Krishna:2022}; for details on the heterodyned likelihood, our priors, and sampler settings, see Appendix~\ref{app:pe}.

\subsection{Simulated Signals} \label{sec:sim-signals}

\begin{table}
    \centering
    \begin{tabular}{cccccccc}
        $\tilde{m}_1$ [$\msun$] & $\tilde{m}_2$ [$\msun$] & $q$ & O4 SNRs & 3G SNRs & Redshift & $m_1$ [$\msun$] & $m_2$ [$\msun$] \\
        \hline
        0.5             & 0.5             & 1.00 & 10.0 & 314.7 & 0.009  & 0.495 & 0.495 \\
                        &                 &      & 29.8 & 934.1 & 0.028  & 0.486 & 0.486 \\
        0.9             & 0.9             & 1.00 & 7.5  & 235.5 & 0.028  & 0.875 & 0.875 \\
                        &                 &      & 16.0 & 502.4 & 0.059  & 0.850 & 0.850 \\
        0.9             & 0.5             & 0.56 & 12.5 & 392.0 & 0.009  & 0.891 & 0.495 \\
                        &                 &      & 37.1 & 1163.4 & 0.028 & 0.875 & 0.486 \\
        1.4             & 0.5             & 0.36 & 21.2 & 665.5 & 0.009  & 1.387 & 0.495 \\
                        &                 &      & 42.4 & 1331.6 & 0.019 & 1.374 & 0.491 \\
        1.0             & 0.1             & 0.10 & 14.3 & 447.5 & 0.009  & 0.991 & 0.099 \\
        1.4             & 0.1             & 0.07 & 13.7 & 429.0 & 0.009  & 1.387 & 0.099\\
    \end{tabular}
    \caption{\swap{Detector-frame component masses $\tilde{m}_1,\tilde{m}_2$} and the mass ratio $q$ of the simulated signals studied in this work.
    At each row \swap{$(\tilde{m}_1,\tilde{m}_2)$} listed here, we simulated a signal for each of $a_1$ chosen from $\{ 0.6, 0.8, 0.9 \}$ and the tilt angle $\theta_1$ chosen as $0$ (no precession) or $\pi / 2$ (precessing).
    For each set of intrinsic parameters \swap{$(\tilde{m}_1, \tilde{m}_2, a_1, \theta_1)$}, we simulate a source in an O4 and 3G detector network with each of the SNRs shown by varying the distance.
    We include representative redshifts \swap{and source-frame component masses $m_1, m_2$} for each combination of \swap{detector-frame} masses and SNR; the redshifts may change by as much $\pm0.001$ to keep the SNR constant as the spin magnitude and tilt are varied.}
    \label{tab:masses}
\end{table}

We consider a set of simulated gravitational-wave signals from binary black hole mergers involving both sub-solar and super-solar mass components.
In Table~\ref{tab:masses}, we show the pairs of \swap{detector-frame component masses $(\tilde{m}_1, \tilde{m}_2)$ of the signals studied in this work; here, $\tilde{m}_1 \geq \tilde{m}_2$}.
For each of these pairs, we consider mergers where the more massive component has a dimensionless spin magnitude $a_1$ of 0.6, 0.8, and 0.9.
These spins are chosen to be greater than the observed maximum spin for a neutron star of $\ans = \maxspin$ \citep{Hessels:2006ze}.
The spin of the less massive component, $a_2$, is always chosen as zero.
At each of these masses and spins, we also consider different orientations of the spin vector of the more massive component, $\vec{S}_1$, characterized by the ``tilt" zenith angle $\theta_1$ between $\vec{S}_1$ and the orbital angular momentum $\vec{L}$.
When the black hole spins are misaligned with $\vec{L}$ (e.g. $\theta_1 > 0$), all angular momenta in the system precess, driven by inertial frame dragging \citep{Apostolatos:1994mx}.
Here, we consider each of two cases, $\theta_1 = 0$ and $\theta_1 = \pi / 2$.
For each set of intrinsic parameters \swap{$(\tilde{m}_1, \tilde{m}_2, a_1, \theta_1)$}, we vary the luminosity distance of the source, $d_L$, to achieve one of the network signal-to-noise ratios (SNRs) listed for each \swap{$(\tilde{m}_1, \tilde{m}_2)$} in Table~\ref{tab:masses}.
\swap{We also note that the black hole masses in the detector frame are larger than the source-frame component masses $m_1, m_2$ due to cosmological redshift by at most 6\%.}
This results in a total of \nsignals different sources.
Our nearest source is at $d_L = 42$ Mpc (a redshift $z \sim 0.009$ assuming the Planck15 cosmology \citep{Planck:2015fie}), and our furthest is at $d_L = 270.7$ Mpc ($z \sim 0.059$).
Additional source parameters are listed in Table~\ref{tab:source-parameters}.
We simulate each signal in a current-generation network of LIGO-Hanford, LIGO-Livingston, and Virgo at design sensitivities for their fourth observing run (O4)\footnote{Using sensitivity curves from \citep{o4designsensitivity}.}, and a next-generation (XG) network of one Cosmic Explorer at the current site of LIGO-Hanford and the Einstein Telescope at the site of Virgo\footnote{Using sensitivity curves from \citep{3gsensitivity}.}.
We use zero-noise realizations of the detector sensitivities; posteriors estimated in zero-noise will be equivalent to the average posterior estimated in many Gaussian-noise realizations \citep{Nissanke:2009kt, Vallisneri:2011ts, Rodriguez:2013oaa, Pankow:2018qpo}.
For all signals considered, we model the gravitational waveform with the phenomenological spin-precessing frequency-domain model \code{IMRPhenomXP} \citep{Pratten:2020ceb}.
We forego higher-order gravitational-wave modes in our analysis for computational expediency, which are expected to be measurable during the merger and ringdown of systems with large total mass or extreme mass ratios.\footnote{We note that \citep{CalderonBustillo:2020kcg} recently showed that higher-order modes can improve the measurement of distance and inclination angle for binary neutron star mergers, comparable to the sub-solar mass black hole mergers studied in this work.}
Thus, our constraints without higher-order modes are conservative upper limits.
Our choices of component masses complement the parameter space studied by recent LIGO-Virgo-KAGRA (LVK) sub-solar mass compact object searches as well as that of Ref.~\citep{Nitz:2022ltl}.
The most recent LVK search considered source-frame masses $m_1 \in [0.2, 10]~\msun$ and $m_2 \in [0.2, 1]~\msun$ \citep{LIGOScientific:2022hai}; Ref.~\citep{Nitz:2022ltl} considered detector-frame masses $\tilde{m}_1 \in [0.1, 100]~\msun$ plus $\tilde{m}_2 \in [0.01, 1]~\msun$ for $\tilde{m}_1 > 20~\msun$ and $\tilde{m}_2 \in [0.1, 1]~\msun$ for $\tilde{m}_1 < 20~\msun$.
As \code{IMRPhenomXP} is only valid for mass ratios $q = m_2 / m_1 > 1 / 20$ \citep{Pratten:2020fqn, Garcia-Quiros:2020qpx}, we choose our minimum $\tilde{m_2} = 0.1~\msun$ consistent with Ref.~\citep{Nitz:2022ltl} and below the LVK search boundary.

\section{Results}

\subsection{Component masses} \label{sec:masses}

\subsubsection{Current-generation Detectors}
\begin{figure}
    \centering
    \includegraphics[width=\linewidth]{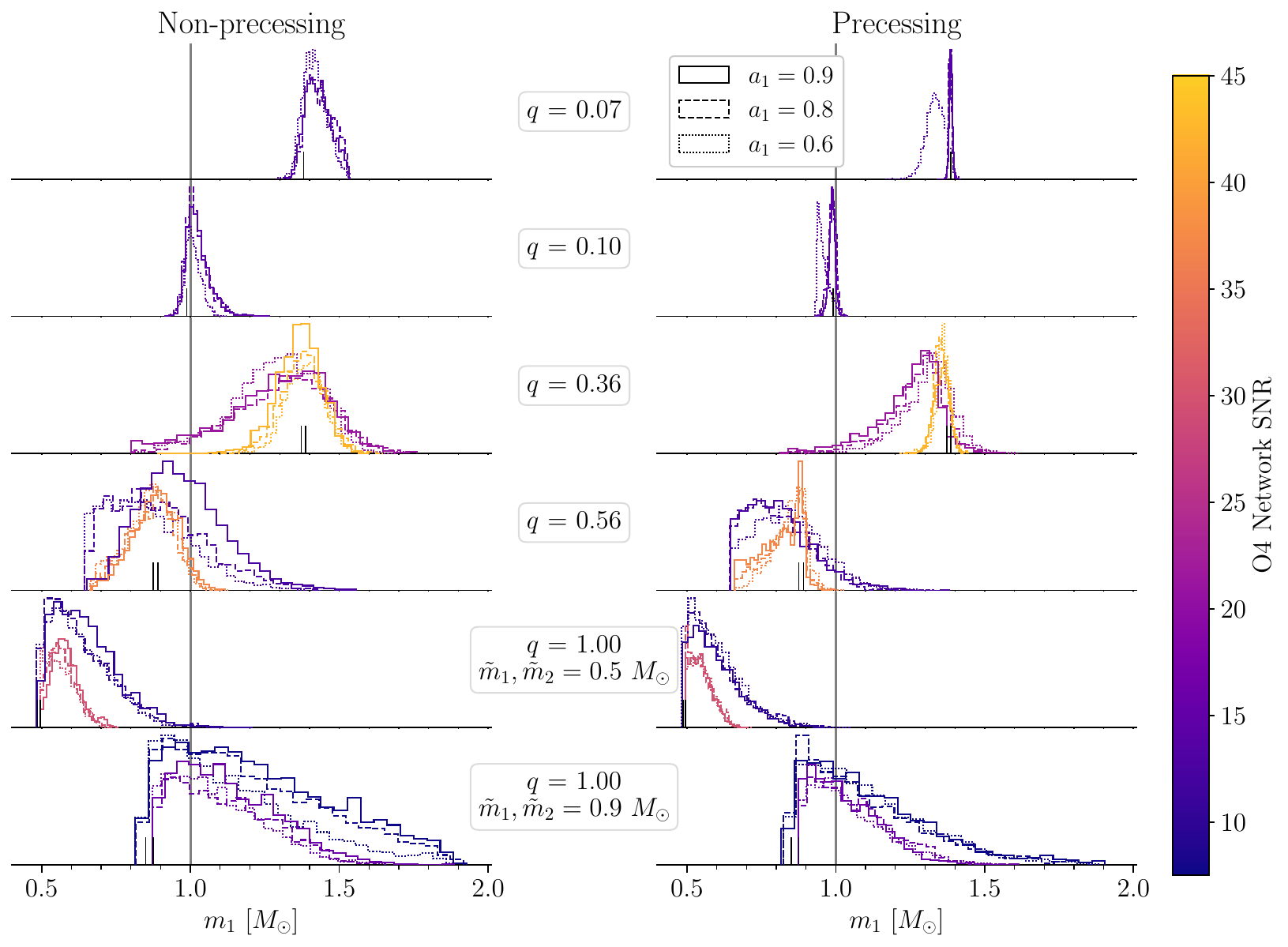}
    \caption{Marginal posterior distributions on the source-frame mass of the heavier black hole, $m_1$ for the simulated signals injected into an O4-design sensitivity network of LIGO-Hanford, LIGO-Livingston, and Virgo.
    Results for non-precessing ($\theta_1 = 0$) sources are shown on the left and those for precessing sources ($\theta_1 = \pi / 2$) on the right.
    The posteriors are colored by the network SNR of the signal.
    The linestyle of the posterior reflects the dimensionless spin magnitude of the more massive black hole, $a_1 = 0.6$ (dotted), 0.8 (dashed), or 0.9 (solid).
    Posteriors are organized by increasing mass ratio of the source, from top to bottom.
    Thin black lines rising from the $m_1$-axis indicate the true value of $m_1$, and a grey line is included at the fiducial mass scale of $1~\msun$. We note that these posteriors are not normalized so that they may be visualized together.
    }
    \label{fig:m1-posteriors}
\end{figure}
Our aim is to determine when we can identify that one or both of the component compact objects has mass $< 1~\msun$.
In general, one expects the estimation of mass parameters to get better and better as the total mass of a binary decreases and the number of inspiral cycles increase; this is why the chirp mass of binary neutron stars is measured much more precisely than that of binary black holes~\cite{Arun:2004hn, Farr:2015lna, Vitale:2016avz} (See also Tab.~IV of Ref.~\cite{LIGOScientific:2021djp}).
In Figure~\ref{fig:m1-posteriors}, we show marginal posteriors on the source-frame mass $m_1$ for the simulated signals described in Section~\ref{sec:sim-signals}, sorted by mass ratio and whether or not the system is precessing.
Posteriors are also colored by network SNR; as expected, as SNR goes from lower values (cooler tones) to higher values (warmer tones), the width of the posteriors decreases.
In all cases, we recover the simulated value (thin black lines) and do so more confidently with decreasing $q$.
This is simply due to the greater distinguishability between $m_1$ and $m_2$ at lower mass ratios.
\swap{We achieve the best measurement of $m_1$ for the precessing, $q = 0.07$ source with $a_1 = 0.9$ with a 90\% credible interval of $1.7 \times 10^{-2}~\msun$.}
\swap{The worst measurement occurs with the $q = 1$, $a_1 = 0.9$, non-precessing source with an SNR of 7.5, with a 90\% credible interval of $0.84~\msun$.}
In general, the network SNR drives the measurement of $m_1$, over $a_1$, however, we do observe that the width of the credible interval weakly depends on $a_1$ at the ${\sim}10\%$ level for non-precessing sources.
In the most extreme example, the 90\% credible interval for the \swap{non-precessing $q = 1$ source at an SNR of 7.5 decreases from $0.84~\msun$ to $0.73~\msun$ as $a_1$ decreases from 0.9 to 0.6 (whereas for the precessing system with the same mass parameters, the credible interval only shrinks from $0.68~\msun$ to $0.61~\msun$)}.

We can also quantify the ``efficiency" of measuring $m_1$, which we evaluate with the ratio between the width of the 90\% credible interval and the SNR; we call this ratio $\alpha$.
As for the width of the credible interval, we find that this ratio typically changes at most at the ${\sim}10\%$ level as we vary $a_1$ and keep SNR the same.
It varies noticeably as the SNR changes, indicating a nonlinear improvement in our measurement of $m_1$ at increasing SNR.
For example, for non-precessing, $q = 0.36$ sources, $\alpha$ improves from \swap{${\sim}1.5 \times 10^{-2}~\msun$ to ${\sim}1.8 \times 10^{-3}~\msun$ as the SNR roughly doubles from 21.2 to 42.4}.

Comparing each posterior to the vertical grey line at $1~\msun$, we are not able to confidently exclude $m_1 \geq 1~\msun$ for sources with $m_1 < 1~\msun$ (bottom three rows).
Among sources with $q = 0.56$, the weakest exclusion of a super-solar mass object occurs for the \swap{non-precessing source with $a_1 = 0.9$, at an SNR of 12.5, where $1~\msun$ occurs at the 60.9\% percentile of the marginal posterior distribution}.
At $q = 1$, with \swap{$\tilde{m}_1 = \tilde{m}_2 = 0.9~\msun$} this exclusion is weaker, with $1~\msun$ occurring between the \swap{23.3\% ($a_1 = 0.9$, non-precessing source with an SNR of 7.5) and 47.0\% ($a_1 = 0.9$, precessing source with an SNR of 16) percentiles}.

\begin{figure}
    \centering
    \includegraphics[width=\linewidth]{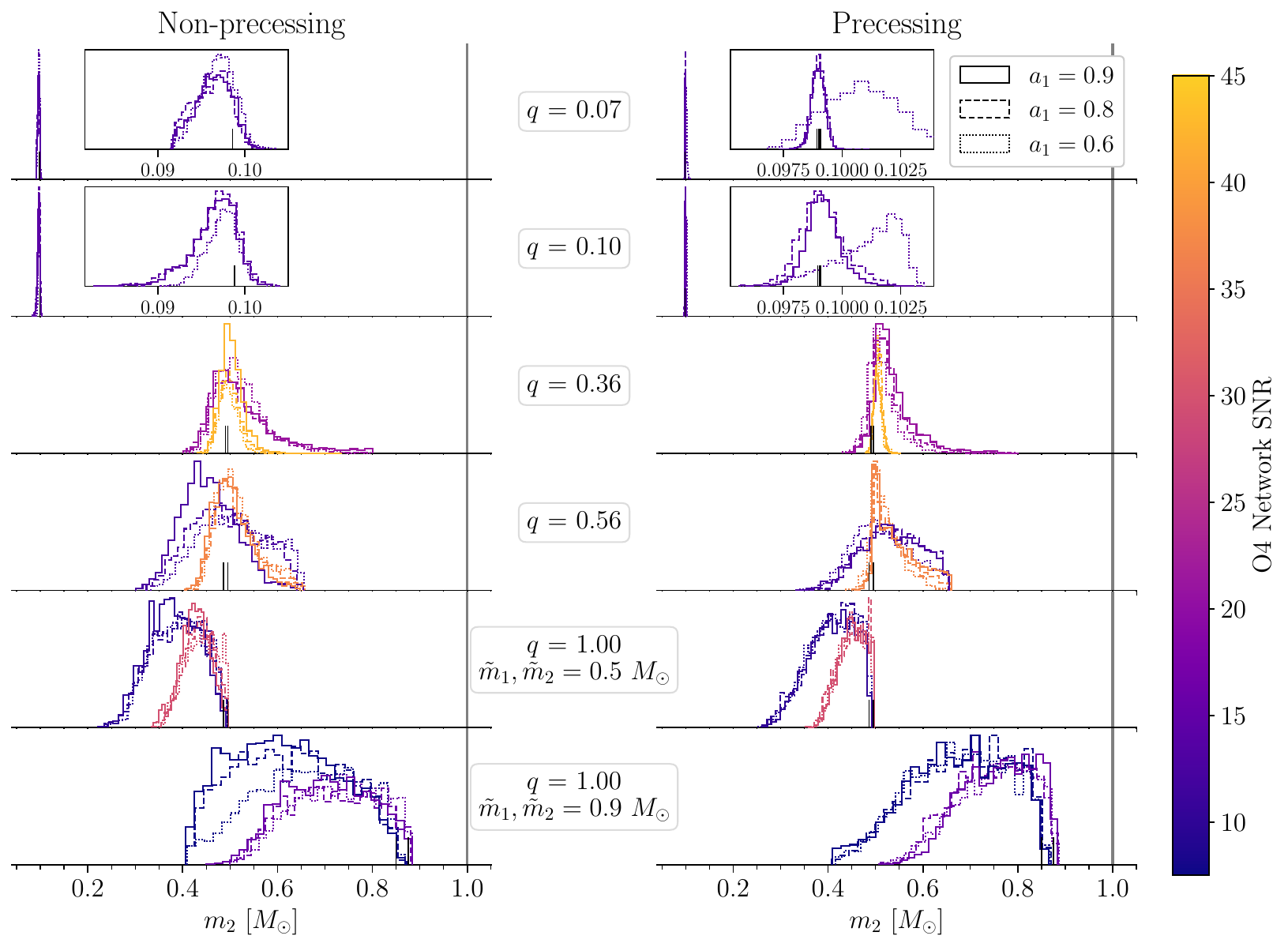}
    \caption{Marginal posterior distributions on the source-frame mass of the lighter black hole, $m_2$ for the simulated signals injected into an O4-design sensitivity network of LIGO-Hanford, LIGO-Livingston, and Virgo.
    Results for non-precessing ($\theta_1 = 0$) sources are shown on the left and those for precessing ($\theta_1 = \pi / 2$) on the right.
    The posteriors are colored by the network SNR of the signal in an O4-design sensitivity network of LIGO-Hanford, LIGO-Livingston, and Virgo.
    The linestyle of the posterior reflects the dimensionless spin magnitude of the more massive black hole, $a_1 = 0.6$ (dotted), 0.8 (dashed), or 0.9 (solid).
    Posteriors are organized by increasing mass ratio of the source, from top to bottom.
    Thin black lines rising from the $m_2$-axis indicate the true value of $m_2$, and a grey line is included at the fiducial mass scale of $1~\msun$.
    We note that these posteriors are not normalized so that they may be visualized together.
    We observe that $m_2 \geq 1~\msun$ is excluded for every simulated signal.}
    \label{fig:m2-posteriors}
\end{figure}

Turning to Figure~\ref{fig:m2-posteriors}, we show marginal posteriors on the source-frame mass $m_2$.
This figure is organized in the same manner as Figure~\ref{fig:m1-posteriors}.
The relationships between the width of the credible intervals, $q$, and SNR are qualitatively much the same as in $m_1$.
However, these posteriors are noticeably more narrow than those for $m_1$ (so much so that for the lowest mass ratios, we provide insets to resolve detail in the histograms).
In particular, our best (worst) measurement occurs for the precessing (non-precessing) source with \swap{$q = 0.07, a_1 = 0.8$ ($q = 1$, \swap{$\tilde{m}_1 = \tilde{m}_2 = 0.9~\msun$}, $a_1 = 0.8$, and network SNR of 7.5), achieving a 90\% credible interval of $9.1 \times 10^{-4}~\msun$ ($0.36~\msun$).}
This improvement follows from the definition of the mass ratio; since $q$ is linear in $m_2$ but $\propto 1 / m_1$, at low mass ratios $q$ is more sensitive to $m_2$ than $m_1$.
Importantly, we observe that \textit{every} posterior excludes $m_2 \geq 1~\msun$.
We also stress that the hard edge observed towards $1~\msun$ in the $q = 1$ results are the posteriors railing on the prior constraint that $m_1 \geq m_2$.
So, at least for systems equivalent to the injections studied in this work, we could confidently report the discovery of a sub-solar mass compact object at SNRs as low as 7.5.
For completeness, we show marginal posteriors on the source-frame chirp mass $\mathcal{M}$ and mass ratio $q$ in the O4 network in Appendix~\ref{app:mchirp-q}.

\subsubsection{Next-generation Detectors}
\begin{figure}[hb]
    \centering
    \includegraphics[width=\linewidth]{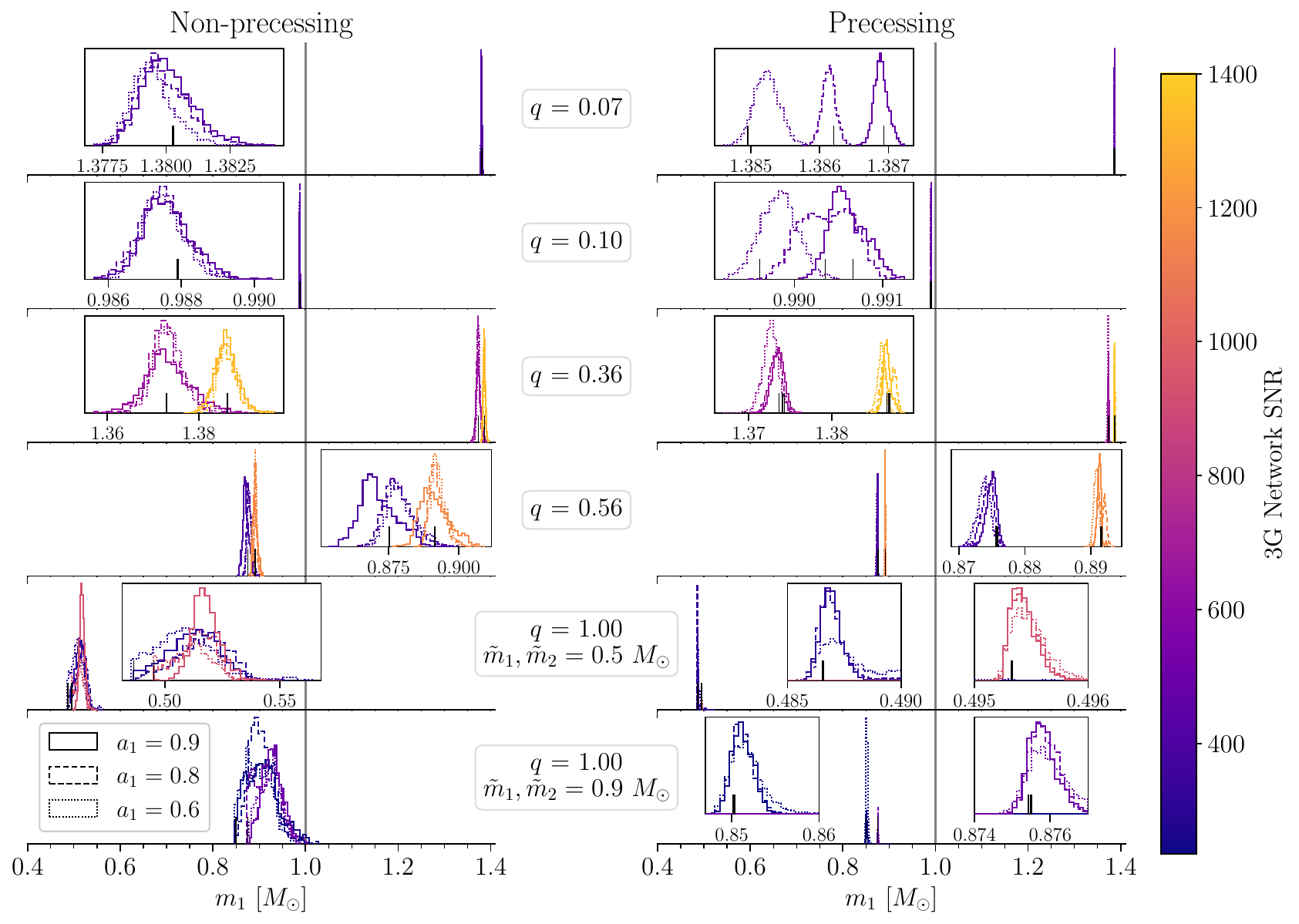}
    \caption{Marginal posterior distributions on the source-frame mass of the heavier black hole, $m_1$ for the simulated signals injected into a network of Cosmic Explorer and the Einstein Telescope.
    Results for non-precessing ($\theta_1 = 0$) sources are shown on the left and those for precessing sources ($\theta_1 = \pi / 2$) on the right.
    The posteriors are colored by the network SNR of the signal.
    The linestyle of the posterior reflects the dimensionless spin magnitude of the more massive black hole, $a_1 = 0.6$ (dotted), 0.8 (dashed), or 0.9 (solid).
    Posteriors are organized by increasing mass ratio of the source, from top to bottom.
    Thin black lines rising from the $m_1$-axis indicate the true value of $m_1$, and a grey line is included at the fiducial mass scale of $1~\msun$.
    We note that these posteriors are not normalized so that they may be visualized together.}
    \label{fig:3g-m1-posteriors}
\end{figure}
\begin{figure}
    \centering
    \includegraphics[width=\linewidth]{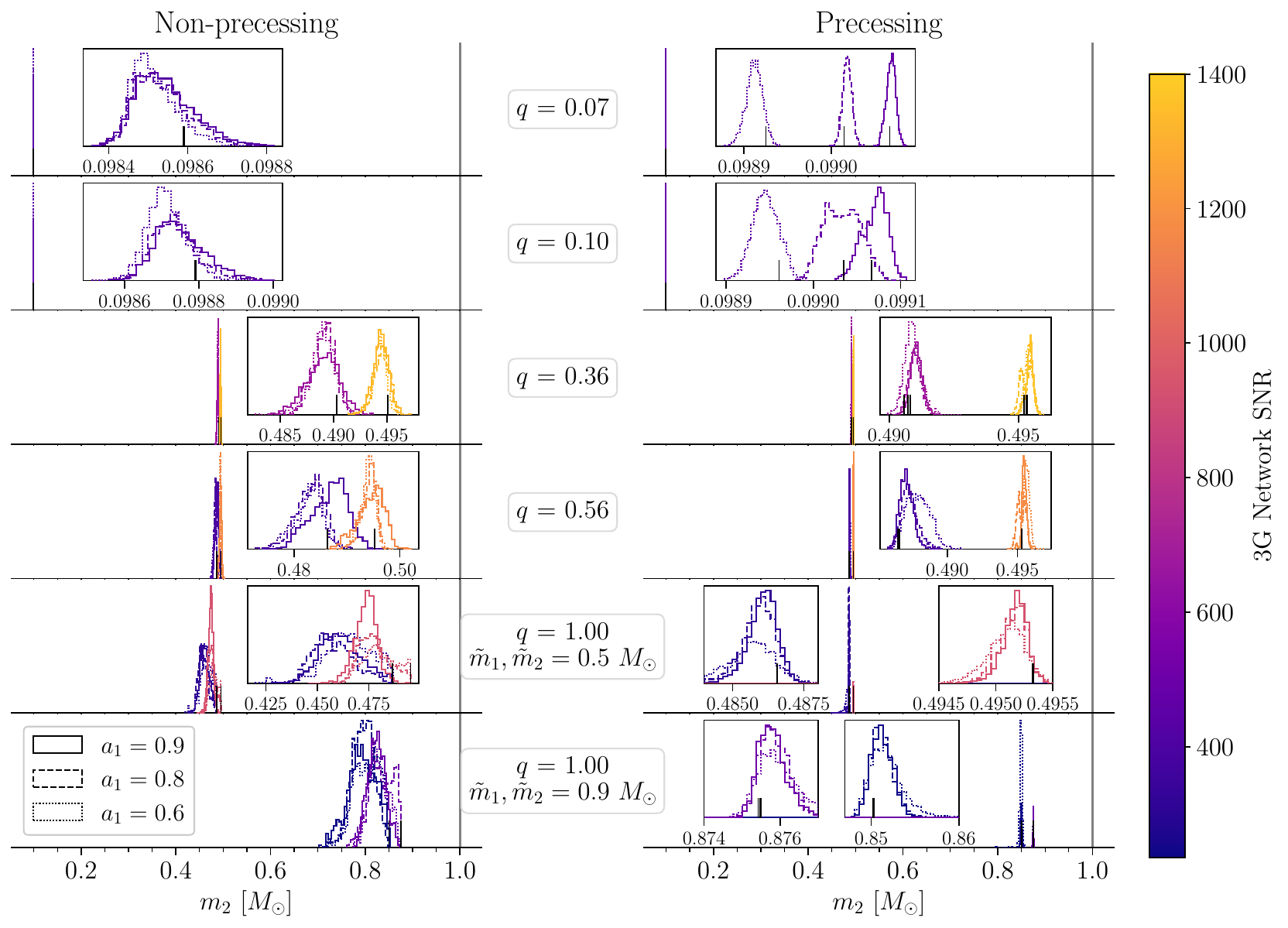}
    \caption{Marginal posterior distributions on the source-frame mass of the lighter black hole, $m_2$ for the simulated signals injected into a network of Cosmic Explorer and the Einstein Telescope.
    Results for non-precessing ($\theta_1 = 0$) sources are shown on the left and those for precessing ($\theta_1 = \pi / 2$) on the right.
    The posteriors are colored by the network SNR of the signal.
    The linestyle of the posterior reflects the dimensionless spin magnitude of the more massive black hole, $a_1 = 0.6$ (dotted), 0.8 (dashed), or 0.9 (solid).
    Posteriors are organized by increasing mass ratio of the source, from top to bottom.
    Thin black lines rising from the $m_2$-axis indicate the true value of $m_2$, and a grey line is included at the fiducial mass scale of $1~\msun$.
    We note that these posteriors are not normalized so that they may be visualized together.}
    \label{fig:3g-m2-posteriors}
\end{figure}
In Figures~\ref{fig:3g-m1-posteriors} and \ref{fig:3g-m2-posteriors}, we show marginal posteriors on the source-frame component masses $m_1$ and $m_2$, respectively, for signals simulated in the XG network of one Cosmic Explorer and Einstein Telescope.
These figures are organized in the same manner as Figures~\ref{fig:m1-posteriors} and \ref{fig:m2-posteriors} in the previous subsection.
For the more massive component, $m_1$, we correctly characterize the compact object as solar, sub-, or super-solar in mass for all the signals studied here, except in the case of non-precessing sources with \swap{detector-frame masses $\tilde{m}_1 = \tilde{m}_2 = 0.9~\msun$}.
\swap{There, $1~\msun$ occurs at worst at the $\gtrsim 99.9\%$ percentile, for the $a_1 = 0.9$ source.}

For the less massive component, $m_2$, we correctly characterize the compact object as sub-solar in nature in all cases, like in current-generation detectors.
Except for signals with $q = 1$, we are also able to constrain both $m_1$ and $m_2$ away from the prior, in particular, without observing any railing on the mass ratio prior.
Again, we stress that the hard edge on the $q = 1$ posteriors is a result of the definition $m_1 \geq m_2$.
We achieve our best (worst) measurement of $m_1$ for the precessing (non-precessing) source with \swap{$q = 0.07$, $a_1 = 0.8$, and a network SNR of 476.0 ($q = 1$,  $\tilde{m}_1 = \tilde{m}_2 = 0.9~\msun$, $a_1 = 0.9$, and SNR of 235.5), that has a 90\% credible of $3.5 \times 10^{-4}~\msun$ ($0.11~\msun$)}.
For $m_2$, we achieve our best and worst measurements with the same sources which have credible intervals of $1.7 \times 10^{-5}~\msun$ ($q = 0.07$, $a_1 = 0.8$, and SNR of 476.0) and $0.10~\msun$ ($q = 1$, $\tilde{m}_1 = \tilde{m}_2 = 0.9~\msun$, $a_1 = 0.9$, and SNR of 235.5), respectively.
This improvement may be driven by the same nonlinear improvement in the measurement efficiency $\alpha$ seen in O4 detectors, which continues in the XG network.
Previously, we saw that for the non-precessing, $q = 0.36$ source, $\alpha$ for $m_1$ is quartered as the SNR doubles; for the same source in a XG network, \swap{$\alpha$ decreases from ${\sim}1.7 \times 10^{-5}~\msun$ to ${\sim}5.8 \times 10^{-6}~\msun$} as the SNR increases from ${\sim}666$ to ${\sim}1333$.
We see a similar pattern in the measurement efficiency of $m_2$.
Overall, we observe an order-of-magnitude improvement in our ability to measure the component masses in XG detectors compared to the O4 network.
We show marginal posteriors on the source-frame chirp mass and mass ratio in the XG network in Appendix~\ref{app:mchirp-q}.

\subsection{Spins}
\subsubsection{Effective Spins in Current-generation Detectors}
\begin{figure}[hb]
    \centering
    \includegraphics[width=\linewidth]{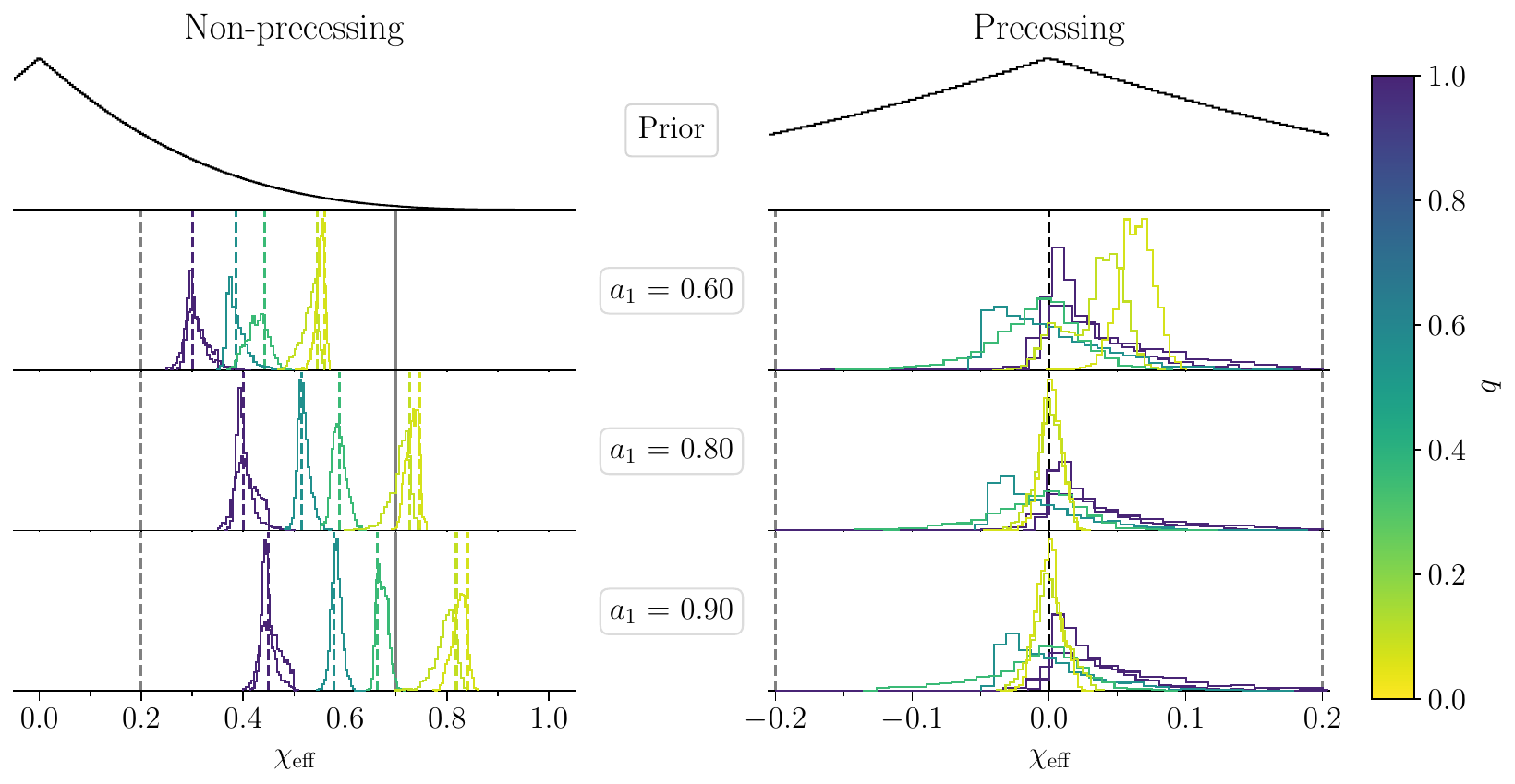}
    \caption{Prior distribution $\pi(\chieff)$ (top row) and marginal posterior distributions on the effective spin \chieff for the quietest signals for a given set of intrinsic parameters $(\tilde{m}_1, \tilde{m}_2, a_1, \theta_1)$, injected into an O4-design sensitivity network of LIGO-Hanford, LIGO-Livingston, and Virgo (remaining rows).
    These are sorted, from top to bottom, by increasing dimensionless spin magnitude on the heavier black hole, $a_1$.
    We color the posterior distributions by the true mass ratio of the source.
    In the left column, we show results for non-precessing ($\theta_1 = 0$) sources, where the truth is indicated with a dashed line colored according to the true $q$.
    In the right column, we show results for precessing ($\theta_1 = \pi / 2$) sources, where the truth is indicated with a dashed black line.
    We also include two critical values of $\chieff$ above which $a_2$ is inconsistent with a neutron star spin, as derived in Appendix~\ref{app:a2-constraints-chieff}.
    These constraints are constructed assuming the spins are aligned (solid grey) and anti-aligned (dashed grey).
    We observe that \chieff is best constrained for non-precessing systems, and at lower mass ratios.
    }
    \label{fig:chieff-posteriors}
\end{figure}
The spin of a compact object may distinguish it from a neutron star.
The maximum observed neutron star spin is that of a pulsar in a binary, with $\ans = \maxspin$ \citep{Hessels:2006ze}.
Whether this value is the theoretical maximal neutron star spin depends on the unknown nuclear equation of state; however, we adopt it as a fiducial value, above which a compact object is inconsistent with a neutron star description.
Even at SNRs as high as ${\sim} 40$, we do not expect to measure the component spin magnitudes or spin tilt angles well on their own \citep{Vitale:2014mka, Purrer:2015nkh, Farr:2015lna}.
However, the leading-order contributions of spins to the gravitational inspiral are measurable and can be parameterized with the effective spin \chieff and effective spin precession \chip.

The effective spin is the mass-weighted projection of the component black hole spins along the orbital angular momentum of the system \citep{Damour:2001tu, Ajith:2009bn, Santamaria:2010yb},
\begin{equation}
    \chieff = \frac{m_1 a_1 \cos \theta_1 + m_2 a_2 \cos \theta_2}{m_1 + m_2} = \frac{a_1 \cos \theta_1 + q a_2 \cos \theta_2}{1 + q}.
\end{equation}

The effective spin \chieff is usually measured better than either component spins, even though it is known to be partially degenerate with the mass ratio for inspiral-dominated (i.e. low mass) systems~\cite{Baird:2012cu, Ng:2018neg}.
When \chieff is zero, the black holes may be non-spinning, or spinning entirely in the plane of the orbit.
When \chieff is +1 (-1), the black holes are spinning parallel to and in the direction of (opposite the direction of) the orbital angular momentum.
In Figure~\ref{fig:chieff-posteriors}, we show marginal posteriors on \chieff for a subset of our runs, sorted by $a_1$ and colored by the mass ratio.
On the left, we show results for non-precessing signals, and on the right, results for precessing signals.
For clarity, we only include the quieter signals at each pair of masses shown in Table~\ref{tab:masses}.
The marginal posteriors on \chieff look nearly identical at different SNRs with all else held equal (see Appendix~\ref{app:extra-spins-posteriors} for the marginal posteriors for the louder events).
Instead, the mass ratio drives the measurement of spin parameters as observed in e.g.~\citep{Vitale:2014mka}.
For non-precessing signals, we confidently recover \chieff at all mass ratios and values of $a_1$.
We find our best result for the source with $q = 0.07$, $a_1 = 0.6$, with a 90\% credible interval of $0.029$, and the worst result for the source with \swap{$q = 1$, $\tilde{m}_1 = \tilde{m}_2 = 0.9~\msun$, $a_1 = 0.9$} with a credible interval of $0.72$.
For precessing signals, we find that $\chieff = 0$ is confidently recovered for all sources, with a better measurement of \chieff at lower $q$.
Quantitatively, the posterior distributions for precessing sources are comparable to their non-precessing counterparts; our best (worst) recovery of \chieff occurs for the precessing source with $q = 0.07$, $a_1 = 0.8$ ($q = 1$, \swap{$\tilde{m}_1 = \tilde{m}_2 = 0.9~\msun$}, $a_1 = 0.9$), yielding a credible interval of 0.025 (0.18).

We also observe that, for some precessing sources with lower spin, we would preferentially report the incorrect \chieff.
In particular, the marginal posterior on \chieff for the precessing, $q = 0.56$ (green-blue color) signal at all spin magnitudes is consistent with zero, but peaks away from zero.
Worse, at $a_1 = 0.6$, the $q = 0.1$ signal exhibits bimodality and the $q = 0.07$ result is inconsistent with $\chieff = 0$.
However, these features are the result of posterior support for $q$ slightly away from its true value, combined with large uncertainties in the spin magnitudes and tilts.
We detail this behavior in Appendix~\ref{app:chieff-q-bias}.

Although it would be easier to directly measure the spin magnitudes as consistent or inconsistent with the maximal neutron star spin, we can indirectly constrain $a_2$ through \chieff if we assume some prior knowledge of the geometry of the system ($a_2$ being the spin of the sub-solar mass object in all of our simulated signals).
In Appendix~\ref{app:a2-constraints-chieff} we find critical values of \chieff for which $a_2 > \ans$, assuming that we know the alignment of the black hole spins.
We include these constraints as grey lines in both columns of Figure~\ref{fig:chieff-posteriors}.
These constraints are constructed with conservative assumptions on the tilts of the black holes, the spin of the more massive black hole, and the mass ratio of the system.
In particular, if $|\chieff|$ is larger than the spin-aligned constraint (solid grey), then the spin of the lighter object (which is a sub-solar mass black hole in all of our simulated sources) \textit{must} be larger than the maximal neutron star spin.
For non-precessing signals, we can exclude $|\chieff|$ below both constraints at the very lowest mass ratios.
Otherwise, we would need to infer that the spins are anti-aligned to exclude spins consistent with a neutron star.
For precessing signals, our inference of \chieff provides no constraint on $a_2$.
This is due to the lossy nature of the \chieff parameterization, as we are trying to recover zero which is degenerate with precessing signals (as we have simulated) as well as systems without any spin.
Since the tilt angles are not well measured, it is difficult to break this degeneracy.
These constraints may be relevant depending on the binary formation mechanism of sub-solar mass black holes and their associated spin distributions.

\begin{figure}
    \centering
    \includegraphics[width=\linewidth]{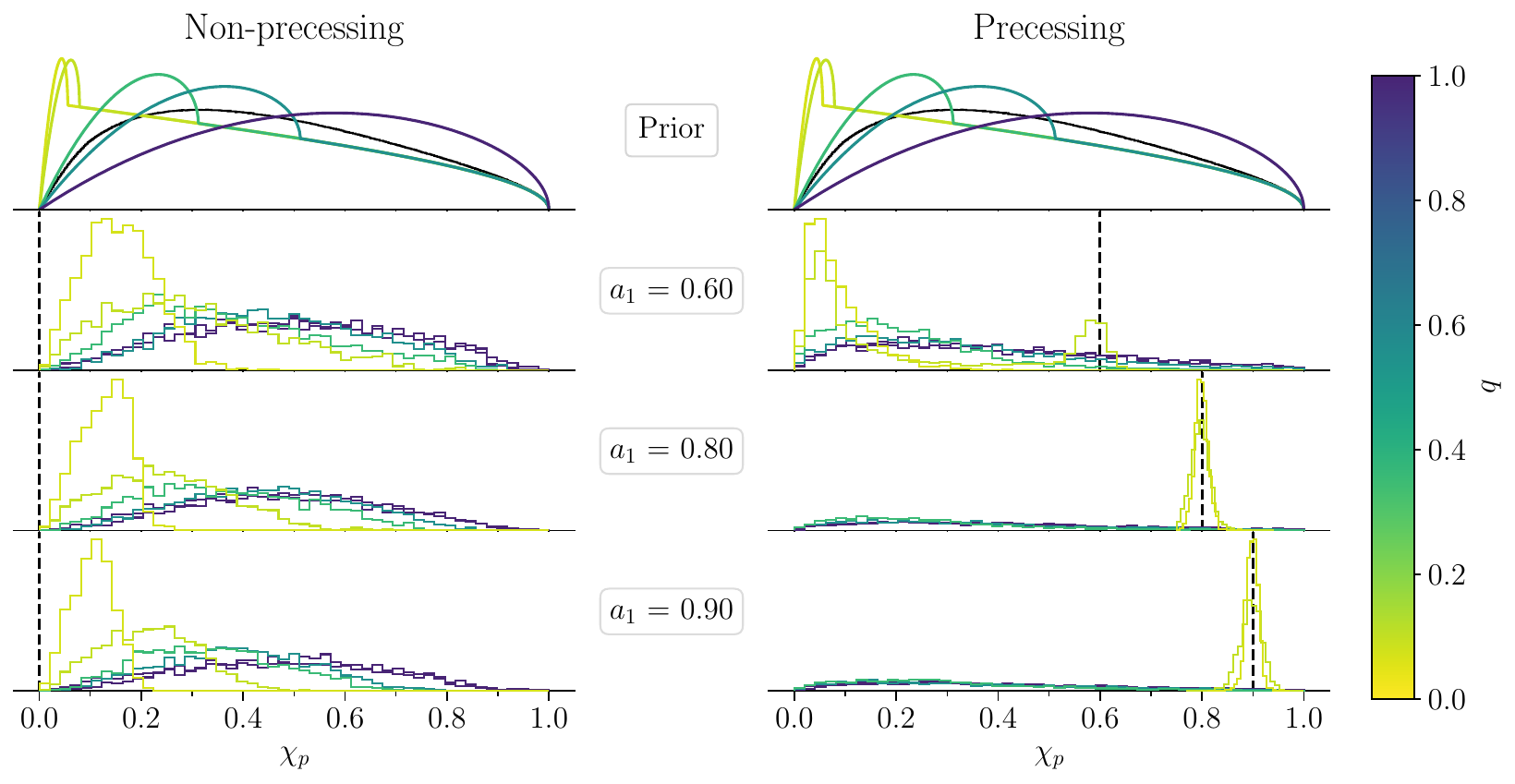}
    \caption{In the top row, we show the prior distribution $\pi(\chip)$ (black) and conditional priors $\pi(\chip|q)$ from \citep{Callister:2021gxf} at the mass ratios listed in Table~\ref{tab:masses}.
    In the remaining rows, we show marginal posterior distributions on \chip for the quietest signals for a given set of intrinsic parameters $(\tilde{m}_1, \tilde{m}_2, a_1, \theta_1)$, injected into an O4-design sensitivity network of LIGO-Hanford, LIGO-Livingston, and Virgo.
    These are sorted, from top to bottom, by increasing dimensionless spin magnitude on the heavier black hole, $a_1$.
    We color the posterior distributions by the true mass ratio of the source.
    In the left column, we show results for non-precessing ($\theta_1 = 0$) sources, and in the right column, we show results for precessing ($\theta_1 = \pi / 2$) sources. 
    The truth is indicated with a dashed black line.
    We observe that \chip is best constrained for precessing systems, and at lower mass ratios.
    }
    \label{fig:chip-posteriors}
\end{figure}

The effective spin precession is the average magnitude of the leading-order contribution to the precession of the orbital angular momentum of the binary, and can be expressed to leading order as \citep{Schmidt:2014iyl},
\begin{equation} \label{eqn:chip}
    \chi_p = \max \left( a_1 \sin \theta_1, q \frac{4q  + 3}{4 + 3 q} a_2 \sin \theta_2 \right).
\end{equation}
While this quantity does not capture all of the relativistic dynamics of precessing binaries (see e.g. \citep{Gerosa:2020aiw}), it best captures precession in systems like those we have simulated, where a single component drives the orbital precession \citep{Schmidt:2014iyl}.
In Figure~\ref{fig:chip-posteriors}, we plot marginal posteriors on $\chi_p$ for the same subset of runs shown in Figure~\ref{fig:chieff-posteriors}, organized in the same manner.
We also include the prior $\pi(\chip)$ and conditional priors $\pi(\chip|q)$, derived in \citep{Callister:2021gxf}.

First, we note that we make a weak (though incorrect) measurement of \chip for non-precessing systems at the most extreme mass ratio we studied, $q = 0.07$.
For each non-precessing $q = 0.07$ source we recovered a posterior without the same ``tail" shown in the priors on \chip in Figure~\ref{fig:chip-posteriors}.
Further, for these sources, the 90\% credible interval on the mass ratio is ${\sim}0.01$, with the true value at the ${\sim}65\%$ percentile; i.e. most of the marginal posterior in $q$ is in smaller mass ratios.
Referring to the conditional priors $\pi(\chip | q)$, we see that lower $q$ tends to increase support near \chip of zero, but instead, we measure $\chip \sim 0.1$.
Thus, the uncertainty on our estimate of $q$ for these sources cannot explain the bias on our measurement of \chip.
For these systems, we may instead be dominated by the uncertainty in the tilt angle, and explaining some of the observed spin in the system with slightly misaligned compact object spins.
For the remaining simulated signals, we can easily compare the posteriors on \chip to $\pi(\chip)$ and $\pi(\chip | q)$, and we find that that we only measure $\chi_p$ for precessing systems at the lowest mass ratios and the highest ($a_1 = 0.8, 0.9$) spins.
This can be explained by remembering that slice of the binary parameter space is where precession will modify the inspiral of the binary the most.

\subsubsection{Component Spin Magnitudes and Tilts in Next-generation Detectors} \label{sec:spins-3g}
\begin{figure}[h]
    \centering
    \includegraphics[width=\linewidth]{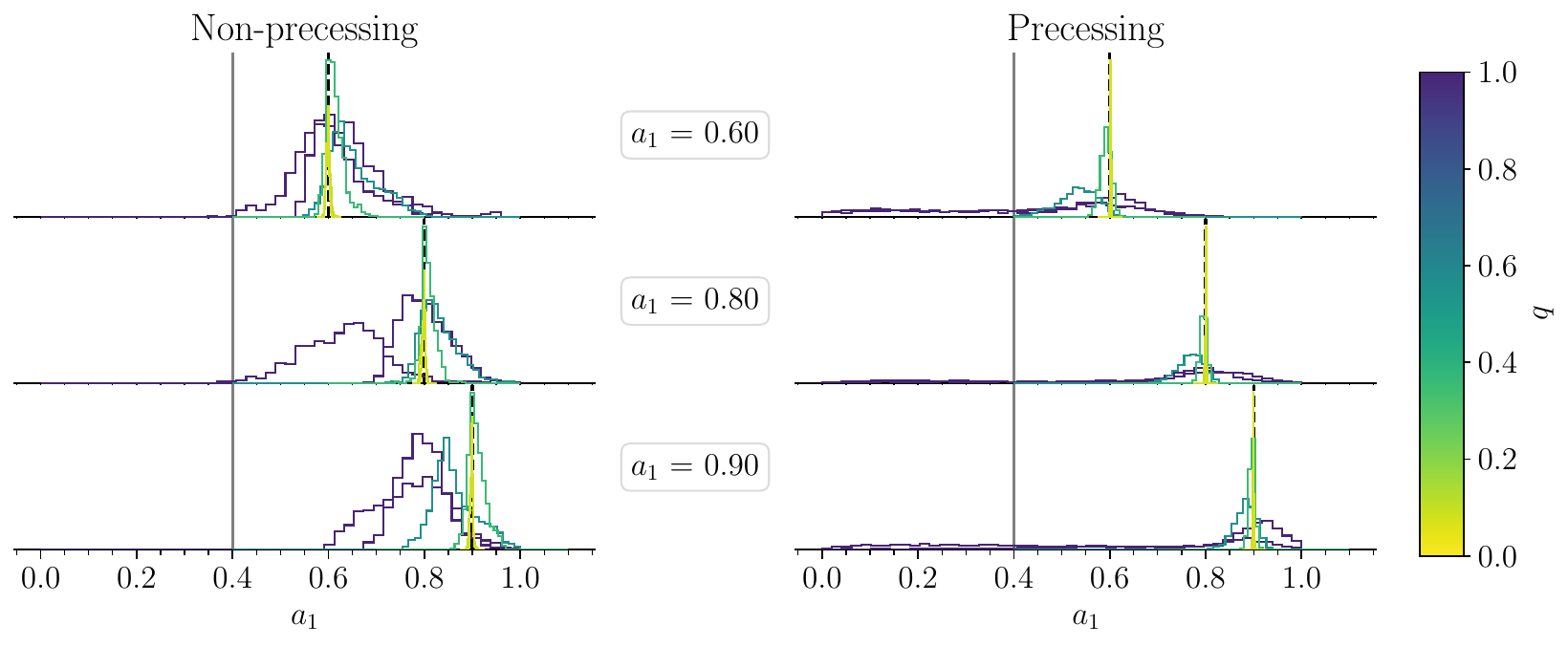}
    \caption{
    Marginal posterior distributions on the dimensionless spin magnitude of the heavier black hole $a_1$, for the quietest signals for a given set of intrinsic parameters $(\tilde{m}_1, \tilde{m}_2, a_1, \theta_1)$, injected into a network of Cosmic Explorer and the Einstein Telescope.
    These are sorted, from top to bottom, by increasing values of $a_1$.
    We color the posterior distributions by the true mass ratio of the source.
    In the left column, we show results for non-precessing ($\theta_1 = 0$) sources, and in the right column, we show results for precessing ($\theta_1 = \pi / 2$) sources. 
    The truth is indicated with a dashed black line.
    For reference, we include a solid grey line at the maximal neutron star spin, $\ans = 0.4$.
    We observe that $a_1$ is measured away from these limits for non-precessing systems and at lower mass ratios.
    }
    \label{fig:3g-a1-posteriors}
\end{figure}

\begin{figure}
    \centering
    \includegraphics[width=\linewidth]{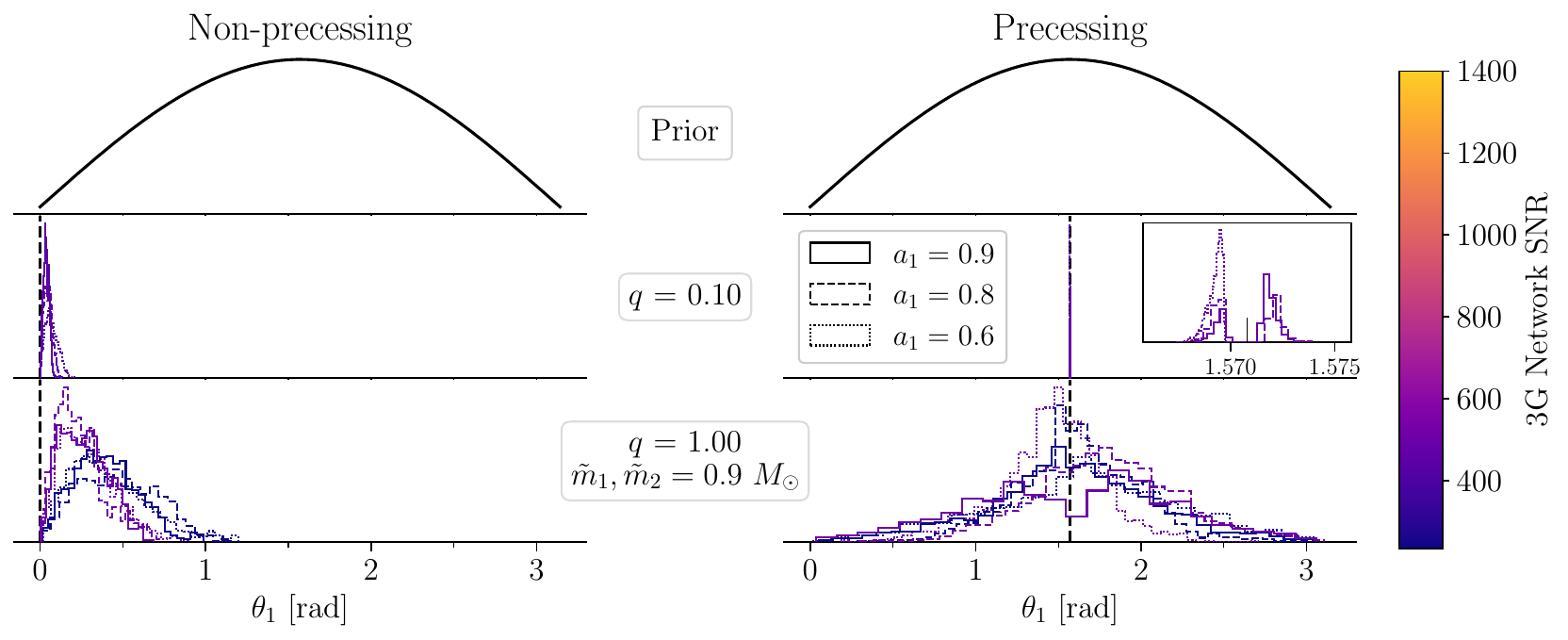}
    \caption{Marginal posterior distributions on the tilt angle between the spin vector of the heavier black hole and the orbital angular momentum, $\theta_1$, for the quietest signals for a given set of intrinsic parameters $(\tilde{m}_1, \tilde{m}_2, a_1, \theta_1)$, injected into a network of Cosmic Explorer and the Einstein Telescope.
    In the top row, we show the prior $\pi(\theta_1)$.
    For brevity, we only include results for sources with $q = 0.10$ (middle row) and $q = 1.0$ (bottom row).
    In the left column, we show results for non-precessing ($\theta_1 = 0$) sources, and in the right column, we show results for precessing ($\theta_1 = \pi / 2$) sources. 
    The truth is indicated with a dashed black line.
    We color the posterior distributions by the network SNR, and their linestyles reflect the true spin magnitude of the heavier black hole, $a_1$.
    We observe that $\theta_1$ is well-measured for non-precessing systems, and at low mass ratios.
    In addition, we note that the lack of posterior support \textit{at} $\theta_1 = \pi / 2$ for precessing, $q = 0.1$ signals likely reflect heterodyning or waveform accuracy issues at large SNRs.
    }
    \label{fig:3g-tilt1-reduced-posteriors}
\end{figure}
In the previous section, we showed that effective spins may be measurable and provide some constraints on the nature of sub-solar mass compact objects in O4 gravitational-wave detectors.
Unfortunately, individual spins cannot be well-constrained at low SNRs.
Of the simulated signals studied in this work, the lowest network SNR achieved in the XG network is \xgsmallestsnr.
These signals are orders of magnitude louder than any gravitational-wave event observed to date and enable us to directly measure the magnitude and angles of binary black hole spin vectors.

In Figure~\ref{fig:3g-a1-posteriors}, we show marginal posteriors on the spin magnitude of the more massive component, $a_1$, for a subset of our signals injected into XG detectors, showing results only for the quieter signals.
The posteriors are colored by the mass ratio of the source, and organized by the true value of $a_1$, increasing from top to bottom.
In all cases, we make a measurement consistent with the true value of $a_1$.
Comparing to the maximal neutron star spin $\ans = 0.4$ (solid grey line), we observe that the best exclusion of such spins occurs for non-precessing systems.
In addition, the constraints on $a_1$ improve with increasing mass ratio.
Among non-precessing systems, we measure $a_1$ the best (worst) for the source with $q = 0.07$, $a_1 = 0.9$ ($q = 1$, \swap{$\tilde{m}_1 = \tilde{m}_2 = 0.9~\msun$}, $a_1 = 0.6$), which yields a 90\% credible interval of $7.7 \times 10^{-3}$ (0.28).
Among precessing systems, we similarly measure $a_1$ the best (worst) for the $q = 0.07$, $a_1 = 0.8$ ($q = 1$, \swap{$\tilde{m}_1 = \tilde{m}_2 = 0.9~\msun$}, $a_1 = 0.8$) source, which yields a 90\% credible interval of $8.2 \times 10^{-4}$ (0.78).
For systems approaching equal masses, we may not be able to exclude $a_1$ consistent with a neutron star if the system is highly precessing with a fully in-plane spin.

In Figure~\ref{fig:3g-tilt1-reduced-posteriors}, we show marginal posteriors on $\theta_1$ for simulated signals with mass ratios $q = 0.1$ and $q = 1$ (only the \swap{$\tilde{m}_1 = \tilde{m}_2 = 0.9~\msun$} runs, for clarity); posteriors at all $q$ can be found in Appendix~\ref{app:extra-spins-posteriors}.
These are colored by network SNR, with a linestyle denoting the value of $a_1$ for each system.
Our analysis employed isotropic priors on $\theta_1$, shown in the top rows of both Figure~\ref{fig:3g-tilt1-reduced-posteriors}.
Comparing the marginal posteriors shown to the priors on $\theta_1$, we find that we can make a measurement of the tilt angle for non-precessing systems, and for precessing systems up until $q = 1$ (for those systems, we make a measurement that excludes a totally aligned or anti-aligned spin with respect to the orbital angular momentum).
For example, for the non-precessing, $q = 0.1$, $a_1 = 0.9$ source, we find a 90\% credible interval of 0.01 radians, and for the non-precessing $q = 1$, \swap{$\tilde{m}_1 = \tilde{m}_2 = 0.9~\msun$}, $a_1 = 0.9$ source with network SNR of 7.5, we find a credible interval of 0.26 radians.
For our precessing sources, the measurement of $\theta_1$ improves at $q = 0.10$ while dramatically broadening at $q = 1.00$; for example, the signal with $q = 0.1$, $a_1 = 0.9$ yields a credible interval of $5.7 \times 10^{-3}$ radians versus an interval of $0.86$ radians for the $q = 1$, $a_1 = 0.9$ source with network SNR of 7.5.
Finally, we note a peculiarity in the posteriors on $\theta_1$ at $q = 0.01$; there, the true value is explicitly excluded, with the distribution peaking to either side of $\pi / 2$.
This is a non-physical effect reflecting a breakdown in the numerical approximation to the prior on the spin components aligned with the orbital angular momentum.
In Appendix~\ref{app:tilt-prior} we demonstrate this prior effect by re-analyzing one simulated signal in the XG network with uniform priors on the tilt angles and magnitudes.
Sampling directly in the spin magnitudes and tilts, we recover a posterior with a similar width as shown in Figure~\ref{fig:3g-tilt1-reduced-posteriors} which peaks at the true $\theta_1$.

While there is no direct constraint on the value of $\theta_1$ for a sub-solar mass black hole mimicker, better measurement of $\theta_1$ improves our measurement of effective spin parameters like \chieff and \chip, which can indirectly constrain the nature of a compact object as discussed for signals in current-generation detectors.
In Appendix~\ref{app:spinprec-thetajn-correlation}, we observe a correlation between \chip and the inclination angle of the system relative to the line of sight which also improves our measurement of binary spin physics.
\FloatBarrier
\subsection{Sky Localization}
\begin{figure}[ht]
    \centering
    \includegraphics[width=0.75\linewidth]{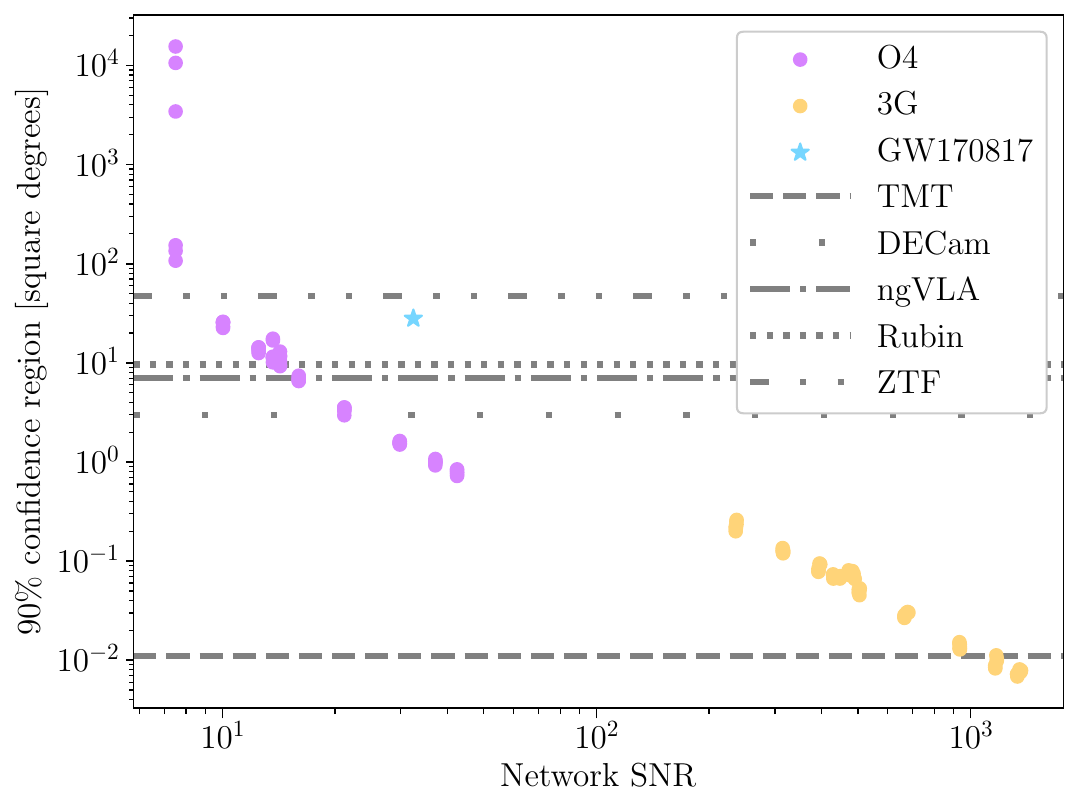}
    \caption{Localization areas i.e. 90\% confidence regions as a function of network SNR for all simulated signals studied in this work.
    Signals injected into an O4-design sensitivity network of LIGO-Hanford, LIGO-Livingston, and Virgo are shown in purple.
    Those injected into a network of Cosmic Explorer and the Einstein Telescope are shown in gold.
    Grey lines indicate the field of view for electromagnetic follow-up observatories, including the Zwicky Transient Facility \citep[ZTF; dash-dot-dotted]{Dekany:2020tyb}, Vera C. Rubin Observatory \citep[dotted]{LSST:2008ijt}, next-generation Very Large Array \citep[ngVLA; dash-dotted]{2018ASPC..517...15S}, Dark Energy Camera \citep[DECam; wide-spaced dots]{DES:2015wtr}, and the Thirty Meter Telescope \citep[TMT; dashed]{2018arXiv180602481S}}
    \label{fig:skyloc-summary}
\end{figure}
Binary neutron star mergers and some neutron star-black hole mergers can produce an electromagnetic (EM) signal counterpart to gravitational-wave emission \citep{Biscoveanu:2022iue}.
While the lack of an EM-counterpart does not strictly rule out a neutron star component, it can reduce the allowed binary parameter space.
In particular, a neutron star in a low-mass ratio merger is expected to be tidally disrupted \citep{Foucart:2018rjc}, and we expect to measure $q$ particularly well for the low-mass ratio systems in this work (c.f. Section~\ref{sec:masses}).
So, if a source is well-localized and no counterpart is observed, a gravitational wave event may become easier to explain with a sub-solar mass black hole component.

In Figure~\ref{fig:skyloc-summary}, we show the area on the sky to which each of our simulated signals can be localized as a function of their network SNRs.
This localization area is the solid angle subtended by the 90\% confidence contour on the marginal joint posterior for the right ascension and declination, computed using \code{ligo.skymap} \citep{Singer:2015ema}.
For comparison, we include two reference scales: the area enclosed by the 90\% confidence contour for the binary neutron star merger GW170817 \citep{gw170817_properties} at a network SNR of ${\sim}32$ (blue star), and the field of view of current- and next-generation optical, infrared, and microwave telescopes that could contribute to the electromagnetic follow-up of a sub-solar mass compact object merger (grey, various linestyles).
We find that sub-solar mass compact object mergers achieve a higher sky localization ``efficiency" than GW170817, in the sense that at a similar network SNR it is possible to localize these sub-solar mass mergers to a region on the sky that is an order of magnitude smaller (similar to the measurement efficiency $\alpha$ introduced in Section~\ref{sec:masses}).
This follows from the fact that longer-duration signals are easier to localize and that all of our simulated systems (with either a lower total mass or mass ratio) will merge much more slowly than a binary neutron star merger like GW170817.
In addition, we observe that the localization for every event, except those at the edge of detectability (at O4 network SNRs of 7.5), is within the field of view for at least one telescope, and most are within the field of view for multiple.
In both scenarios (O4 and XG), we see that a compact object merger involving a sub-solar mass component will be very well-localized compared to our electromagnetic follow-up capabilities, reducing the chance of missing an associated electromagnetic counterpart if something other than black holes is involved in the merger.
We also note that our analysis did not include higher-order effects like the time-evolution of the detector antenna pattern due to the rotation of the Earth or the finite size of the detector relative to the wave; however, we expect these to further improve the localization of long-duration signals \citep{Baral:2023xst}.

\section{Discussion}

Here, we have studied the measurability of the masses, spins, and sky location 
for a set of spinning, precessing and non-precessing, quasi-circular binary black hole mergers involving at least one sub-solar mass component.
Using a spin-precessing waveform and heterodyned likelihood, we performed parameter estimation on these signals injected into a network of LIGO-Hanford, LIGO-Livingston, and Virgo at O4-design sensitivity, and a network of next-generation detectors, Cosmic Explorer and the Einstein Telescope.

% discuss masses
We found that the long duration (${\sim}$1000's of seconds) of these signals engendered precise measurement of the black hole masses.
We confidently identify a sub-solar mass compact object in all of the signals we studied, \textit{at current design sensitivities}, down to network signal-to-noise ratios at the threshold of detectability.
In particular, we confidently excluded the least massive component as being $\geq 1~\msun$ in mass even for an equal-mass 0.9-0.9\msun merger with an SNR of 7.5, corresponding to a luminosity distance of ${\sim} 125$ Mpc, or roughly three times the distance of the binary neutron star merger GW170817.
We observed that the SNR drives the measurement of the component masses over the spins, and noted an increasing, nonlinear relationship between the width of the 90\% credible interval on $m_1$ or $m_2$ and the SNR.
Driven by this improvement in measurement efficiency at high SNRs, we found that next-generation detectors will enable exquisite measurements of the source-frame component masses towards the $10^{-5}~\msun$ level, enabling confident measurement of the compact objects we studied as super-, sub-, or solar in mass.

% discuss spins
We then looked at the spins.
In current generation detectors, we do not expect SNRs high enough to often confidently measure the component spin magnitudes nor their tilt angles with respect to the orbital angular moment.
However, we found that the leading-order effective spin parameters \chieff and \chip can be well-measured for non-precessing and precessing sources, respectively, in the O4 network.
With strong assumptions on the binary geometry, we found that for non-precessing signals with the lowest mass ratios ($q = 0.07, 0.10$), we could use \chieff to confidently exclude spins consistent with the fastest spinning neutron stars.
In next-generation detectors, signals will be orders-of-magnitude louder at or beyond distances comparable to GW170817.
This enabled us to directly measure the spin magnitude and tilt of the heavier component, and exclude spins consistent with theoretical maximal neutron star spins for non-precessing systems, and precessing systems at low mass ratios.

% discuss localization
Binary neutron star and neutron star-black hole mergers may produce an electromagnetic counterpart; if none is found, especially at low mass ratios, that may rule out a neutron star component in a binary merger.
Again enabled by the long duration of these signals, we found that nearly all of the sub-solar mass sources we studied in the three-detector, O4-design sensitivity network would be localized within the field of view of at least one current- or next-generation electromagnetic follow-up instrument.
We noted that many of these sources are more ``efficiently" localized than GW170817, achieving smaller 90\% confidence regions on the sky at similar SNRs.
All of the sources in the two-detector, next-generation gravitational wave network were localized within the field of view of four current- and next-generation electromagnetic follow-up instruments.

% tidal deformability + future work
At least among the sources studied here, current-gravitational wave detectors would be able to confidently report the discovery of a sub-solar mass component of a compact object merger plus some measurement of the binary spin geometry and sky location, enabling unique constraints on the properties of dark matter and the physics of the early Universe.
However, the ``smoking gun" that characterizes a compact object is tidal deformability, which measures how the material of a compact object responds to the gravitational field of its companion.
At present, there are no waveform models including tidal deformability that are calibrated for the low mass ratios, or small component masses studied in this work.
Recently, \citep{Markin:2023fxx} performed the first-ever numerical relativity simulation of a neutron star-sub-solar mass black hole merger.
They found significant dephasing between the waveform predicted by numerical relativity and phenomenological or surrogate waveform predictions for the same merger.
Once waveform models involving tides are developed for sub-solar mass compact object mergers, future work like ours could explore the measurability of those tides.

\section{Acknowledgements}

We thank Katerina Chatziioannou, Tim Detrich, and Carl Johan-Haster for enlightening discussions around the measurement and modeling of tidal deformability in neutron star/sub-solar mass black hole mergers, Geraint Pratten for assistance in understanding the \code{IMRPhenomX} family of waveforms, and Leo Singer for guidance on estimating sky localization regions.
We additionally thank Divya Singh, Ester Ruiz-Morales, and Aditya Vijaykumar for their helpful comments on this manuscript.
This work was inspired by the 2020 ``Workshop on Gravitational Wave Constraints on Dark Compact Objects", which was supported by a Gordon and Betty Moore Foundation Fundamental Physics Innovation Convening Award and the American Physical Society.
This material is based upon work supported by NSF's LIGO Laboratory which is a major facility fully funded by the National Science Foundation.
The authors are grateful for computing resources provided by the California Institute of Technology and supported by National Science Foundation Grants PHY-0757058 and PHY-0823459.
SV is partially supported by NSF through the award PHY-2045740.
NEW is supported by the La Gattuta Physics Fund and the Henry W. Kendall (1955) Fellowship Fund.
CT is supported by an MKI Kavli Fellowship.

\textit{Software:} \code{bilby v2} \citep{Ashton:2018jfp, Romero-Shaw:2020owr}, \code{astropy v5.1} \citep{Astropy:2022ucr}, \code{pesummary v0.13.10} \citep{Hoy:2020vys}, \code{healpy v1.16.1} \citep{Gorski:2004by, Zonca2019}, \code{ligo.skymap v1.0.3} \citep{Singer:2015ema}, \code{corner.py v2.2.1} \citep{corner} , \code{numpy v1.22.3} \citep{harris2020array}, \code{scipy v1.8.1} \citep{2020SciPy-NMeth}, \code{pandas v1.4.2} \citep{jeff_reback_2022_6408044, mckinney-proc-scipy-2010}, \code{dynesty v2.1.0} \citep{Speagle:2019ivv}, \code{LALSimulation v4.0.2} \citep{lalsuite, swiglal}

\appendix

\section{Parameter Estimation} \label{app:pe}
\subsection{Waveform Model and Heterodyned Likelihood}

We use the Whittle likelihood approximation in the frequency domain for the residual noise \citep{Romano:2016dpx},
\begin{equation} \label{eqn:likelihood}
    \mathcal{L}(d | \theta) = \prod_{i,j} \frac{1}{2\pi S_{ij}} \exp\left( -\frac{4}{T}\frac{|d_{ij} - h(\theta)_{ij}|^2}{S_{ij}} \right).
\end{equation}
where the products run over the gravitational wave detectors in a network and the frequency bins of the data, $d_{ij}$ is the data, $h(\theta)_{ij}$ is the strain of the astrophysical signal, $S_{ij}$ is the power spectral density, and $T$ is the duration of the data.
We inject simulated signals into zero noise realizations of the detectors, so the power spectral densities are (theoretical) design sensitivity curves.
The strain is evaluated with the frequency-domain waveform model \code{IMRPhenomXP} which includes spin precession effects \citep{Pratten:2020ceb}.
We choose a similar description of spin-precession as used in \code{IMRPhenomPv2} \citep{Khan:2018fmp} by choosing the flag \code{PhenomXPrecVersion = 104} (see Table III, Appendix F of \citep{Pratten:2020ceb}).

In practice, taking the log-likelihood is commonly expressed in terms of a complex-valued inner product between the data and the waveform strain, weighted by the power spectral density (see Appendix B of \citep{2019PASA...36...10T}).
Evaluating this inner product over $\mathcal{O}(10^6)$ (as we have for signals $\mathcal{O}(10^3)$ seconds in duration) is computationally expensive.
A heterodyned likelihood \citep{Cornish:2021lje} (an approximation also known as relative binning \citep{Zackay:2018qdy}) approximates Equation~\ref{eqn:likelihood} by expanding the inner product between the data and the strain about its value at some fiducial values of $\theta$, at a few frequencies whose spacing is chosen to minimize the change in waveform phase between frequencies.
The \code{bilby} implementation of a heterodyned likelihood is described in Ref.~\citep{Krishna:2022}.
Following the description of \citep{Zackay:2018qdy}, the heterodyned likelihood relies on hyperparameters $\epsilon$ and $\chi$ which control the tolerance for this inter-frequency dephasing (see Equations 9 and 10 of \citep{Zackay:2018qdy}).
In this work, we choose $\epsilon = 0.025$ and $\chi = 1$, which translates to an evaluation of the likelihood at only 1242 frequencies.
For each simulated signal we study, we choose the fiducial values to be the true values chosen for each source.

\subsection{Priors}

Binary black hole parameter definitions follow those of Table E1 in \citep{Romero-Shaw:2020owr}.
Conceptually, we adopt the following priors:
\begin{itemize}
    \item \textit{Component masses}: We adopt uniform priors on \swap{the detector-frame masses $\tilde{m}_1, \tilde{m}_2$}; under these assumptions, we sample in the mass ratio $q$ and detector-frame chirp mass $\tilde{\mathcal{M}}$ as these control the leading-order contributions to the gravitational wave strain.
    We additionally constrain these priors by requiring \swap{$\tilde{m}_1, \tilde{m}_2 \leq 2~\msun$} in all analyses.
    In principle, this allows us to recover super-solar component masses while reducing the volume of the parameter space the sampling procedure may explore.
    
    \item \textit{Luminosity distance}: Uniform prior on the coalescence time in the source frame and the enclosed comoving volume, assuming the Planck15 cosmology \citep{Planck:2015fie}.
    
    \item \textit{Spins}: We assume isotropic priors on the spin vectors of the component black holes, parameterized in terms 
    of the components of the spins aligned with the orbital angular momentum $\vec{L}$, $\chi_{1,2}$, the components pointing in the orbital plane $\chi^{\perp}_{1,2}$, the azimuthal angle (taking $\vec{L}$ to point along the $z$-direction) between the spins, $\phi_{12}$, and the azimuthal angle between $\vec{L}$ and the total angular momentum.

    \item \textit{Sky location and orientation}: We assume isotropic priors for the orientation and location of the binary on the sky, parameterized by the time of coalescence, azimuth, and zenith as defined at LIGO-Hanford.
    Since the time at a given detector is what is measured in practice, this parameterization is more efficient to sample in than directly sampling in the right ascension, declination, and time of coalescence as measured at the center of the Earth.
\end{itemize}

In all of our analyses, we choose priors that encompass the resultant posteriors.

\subsection{Nested Sampling in \code{bilby}}
The strategy of nested sampling is to estimate the Bayesian evidence $\mathcal{Z}$ as the integral of the likelihood surface over the prior mass $X$,
\begin{equation}
    \mathcal{Z} = \int_0^1 \mathcal{L}(X) dX \approx \sum_i w_i \mathcal{L}_i
\end{equation}
where $i$ enumerates likelihood contours of value $\mathcal{L}_i$ and the weights $w_i$ are the the prior mass enclosed by those contours.
The likelihood contours are chosen to enclose increasingly narrow regions of the total prior mass; this sorting allows us to estimate $\mathcal{Z}$ without explicitly referring to the complicated, high-dimensional geometry of the likelihood surface \citep{Skilling:2006gxv}.
While evaluating the likelihood is a well-defined computation using Equation~\ref{eqn:likelihood} and given a model of the gravitational waveform, determining the prior mass enclosed by some likelihood iso-contour is non-trivial.
We can estimate $w_i$ with uncertainty e.g. with Markov Chain Monte Carlo (MCMC) chains sampling within some bounds of the prior mass.
Here, we use the nested sampling algorithm implemented by \code{dynesty} \citep{Speagle:2019ivv} and employ Differential Evolution-MCMC (DE-MC), as implemented in \code{bilby}, to estimate $w_i$ \citep{2006S&C....16..239T}.
In Table~\ref{tab:sampler-kwargs}, we record the sampler settings provided to \code{bilby} when we conduct parameter estimation with nested sampling for our simulated signals.

\begin{table}[hb]
    \centering
    \begin{tabular}{|c|c|}
        \hline
        Sampler Argument & Value \\
        \hline
        \code{nlive} & 500 \\
        \code{walks} & 100 \\
        \code{naccept} & 60 \\
        \code{sample} & \code{`acceptance-walk'} \\
        \code{proposals} & \code{[`diff']} \\
        \code{bound} & \code{`live'} \\
        \hline
    \end{tabular}
    \caption{Sampler arguments used in our analysis of simulated signals as defined for the \code{bilby} implementation of \code{dynesty} and Differential Evolution-MCMC. We note that a few runs use \code{nlive = 2000} to achieve convergence.
    }
    \label{tab:sampler-kwargs}
\end{table}

\begin{table}[hb]
    \centering
    \begin{tabular}{|c|c|c|}
        \hline
        Parameter & Units & Value \\
        \hline
        \multicolumn{3}{|c|}{Intrinsic Parameters} \\
        \hline
        $m_1, m_2$ & \msun & See Table~\ref{tab:masses} \\

        $a_1, a_2$ & -- & $a_1$ is one of $\{ 0.6, 0.8, 0.9 \}$ \\
                   &    & $a_2 = 0$ \\
        $\theta_1, \theta_2$ & radians & $\theta_1$ is one of $\{0, \pi / 2$\} \\ 
                             &         & $\theta_2 = 0$ \\
        $\phi_{12}$ & radians & 1.7 \\
        $\phi_{jl}$ & radians & 0.3 \\
        \hline
        \multicolumn{3}{|c|}{Extrinsic Parameters} \\
        \hline
        $d_L$ & Mpc & Changed to achieve the SNRs in Table~\ref{tab:masses}. \\
        \thetajn & radians & 0.4 \\
        Right ascension & radians & 1.375 \\
        Declination & radians & -1.2108 \\
        $\psi$ & radians & 2.659 \\
        $t_c$ & GPS Time & 1126259642.413 \\
        $\phi_c$ & seconds & 1.3 \\
        
        \hline
    \end{tabular}
    \caption{Complete set of binary black hole source parameters for the simulated signals studied in this work, including the intrinsic parameters of the black holes and the extrinsic parameters of the binary's location and orientation.
    Parameters of the angular momenta are evaluated at a reference frequency of 20 Hz.
    Parameter definitions follow Table~E1 of \citep{Romero-Shaw:2020owr}.}
    \label{tab:source-parameters}
\end{table}

\pagebreak
\clearpage

\section{Marginal Posteriors on Chirp Mass and Mass Ratio} \label{app:mchirp-q}
\begin{figure}[ht]
    \centering
    \includegraphics[width=\linewidth]{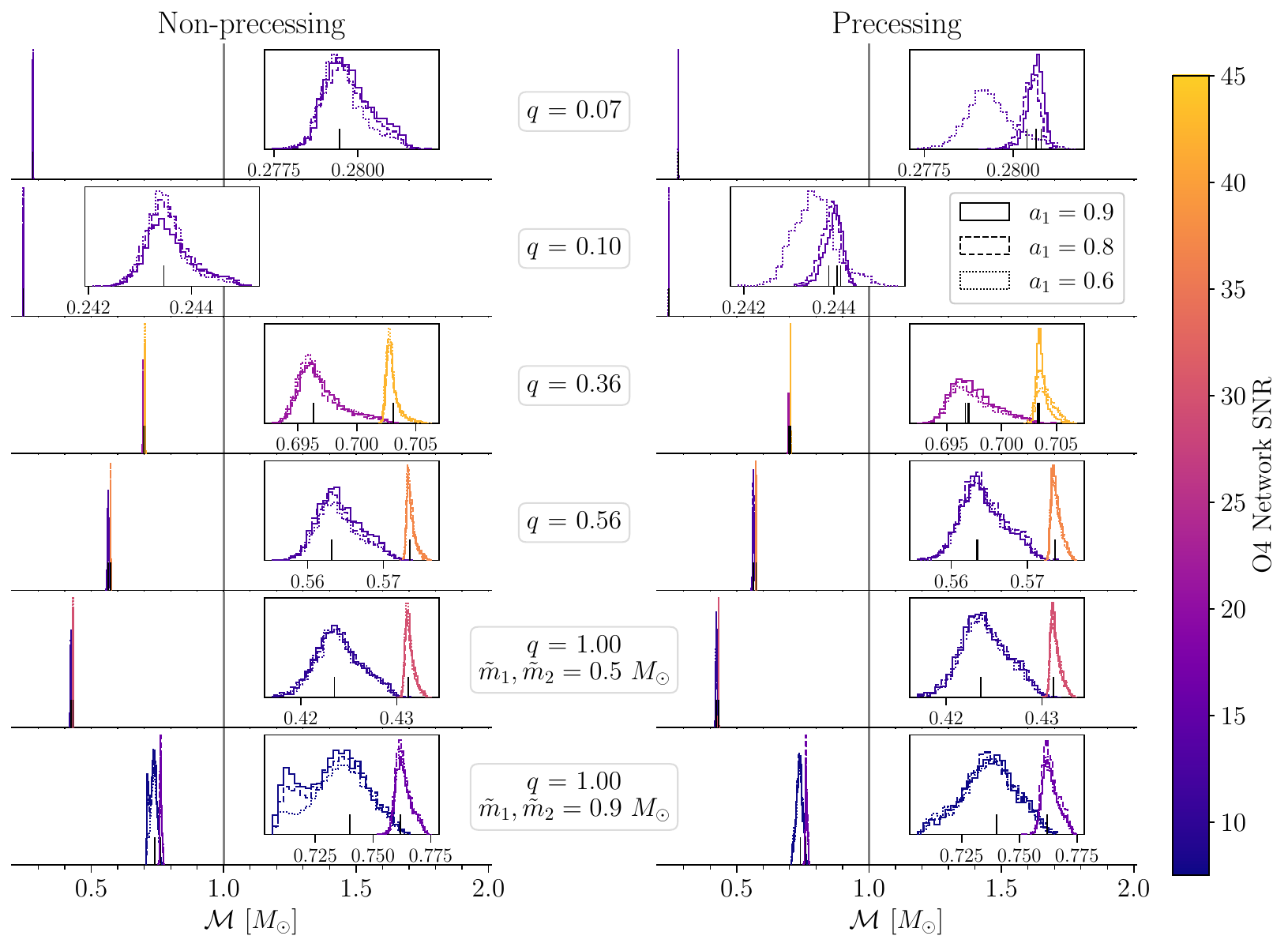}
    \caption{
    Marginal posterior distributions on the source-frame chirp mass $\mathcal{M}$ for the simulated signals injected into an O4-design sensitivity network of LIGO-Hanford, LIGO-Livingston, and Virgo.
    Results for non-precessing ($\theta_1 = 0$) sources are shown on the left and those for precessing sources ($\theta_1 = \pi / 2$) on the right.
    The posteriors are colored by the network SNR of the signal.
    The linestyle of the posterior reflects the dimensionless spin magnitude of the more massive black hole, $a_1 = 0.6$ (dotted), 0.8 (dashed), or 0.9 (solid).
    Posteriors are organized by increasing mass ratio of the source, from top to bottom.
    Thin black lines rising from the $\mathcal{M}$-axis indicate the true value of $\mathcal{M}$, and a grey line is included at the fiducial mass scale of $1~\msun$.
    We note that these posteriors are not normalized so that they may be visualized together.}
    \label{fig:o4-mchirp}
\end{figure}
\begin{figure}[ht]
    \centering
    \includegraphics[width=\linewidth]{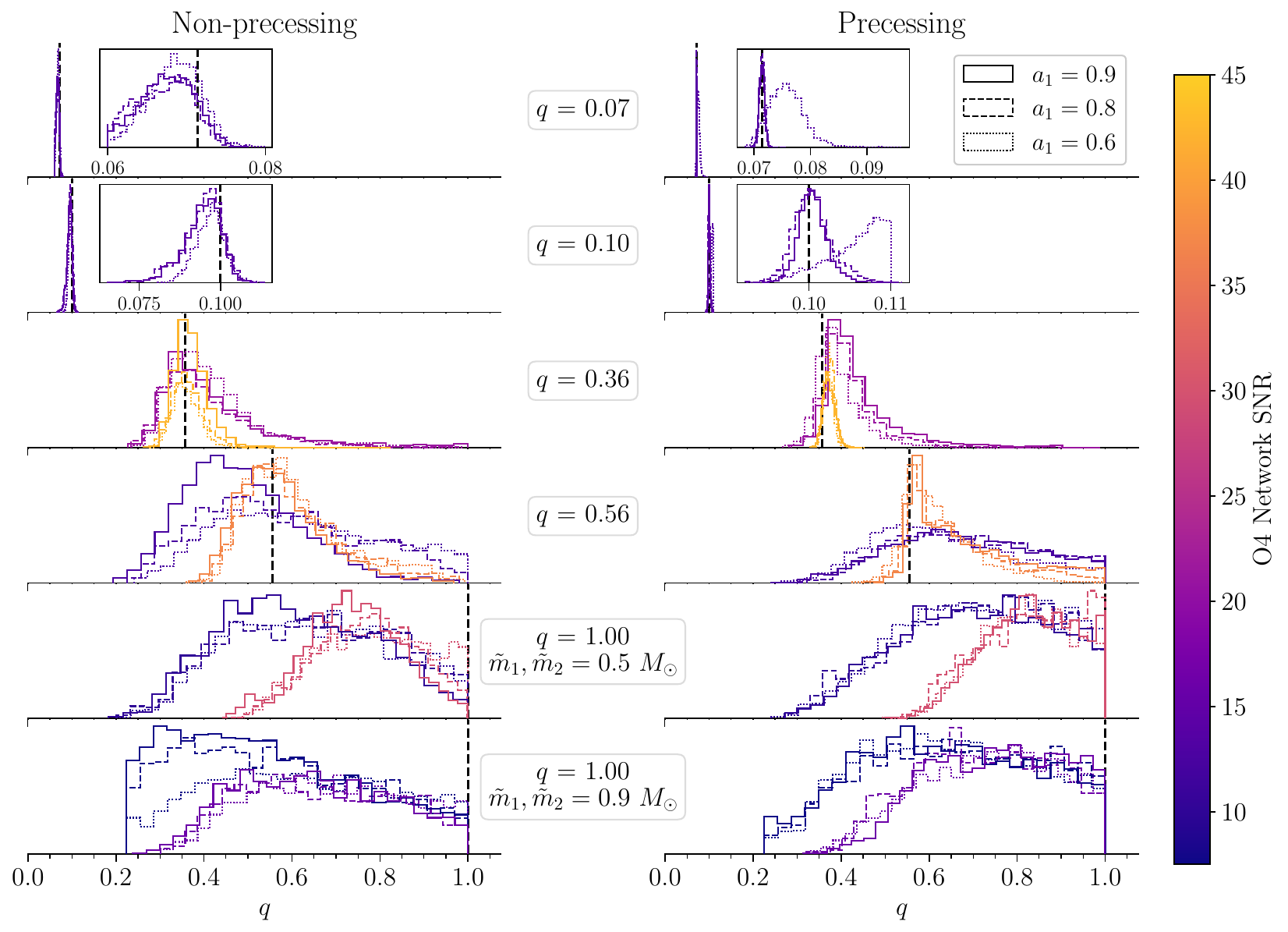}
    \caption
    {Marginal posterior distributions on the mass ratio $q$ for the simulated signals injected into an O4-design sensitivity network of LIGO-Hanford, LIGO-Livingston, and Virgo.
    Results for non-precessing ($\theta_1 = 0$) sources are shown on the left and those for precessing sources ($\theta_1 = \pi / 2$) on the right.
    The posteriors are colored by the network SNR of the signal.
    The linestyle of the posterior reflects the dimensionless spin magnitude of the more massive black hole, $a_1 = 0.6$ (dotted), 0.8 (dashed), or 0.9 (solid).
    Posteriors are organized by increasing mass ratio of the source, from top to bottom.
    Dashed black lines indicate the true value of $q$.
    We note that these posteriors are not normalized so that they may be visualized together.}
    \label{fig:o4-massratio}
\end{figure}
\begin{figure}[ht]
    \centering
    \includegraphics[width=\linewidth]{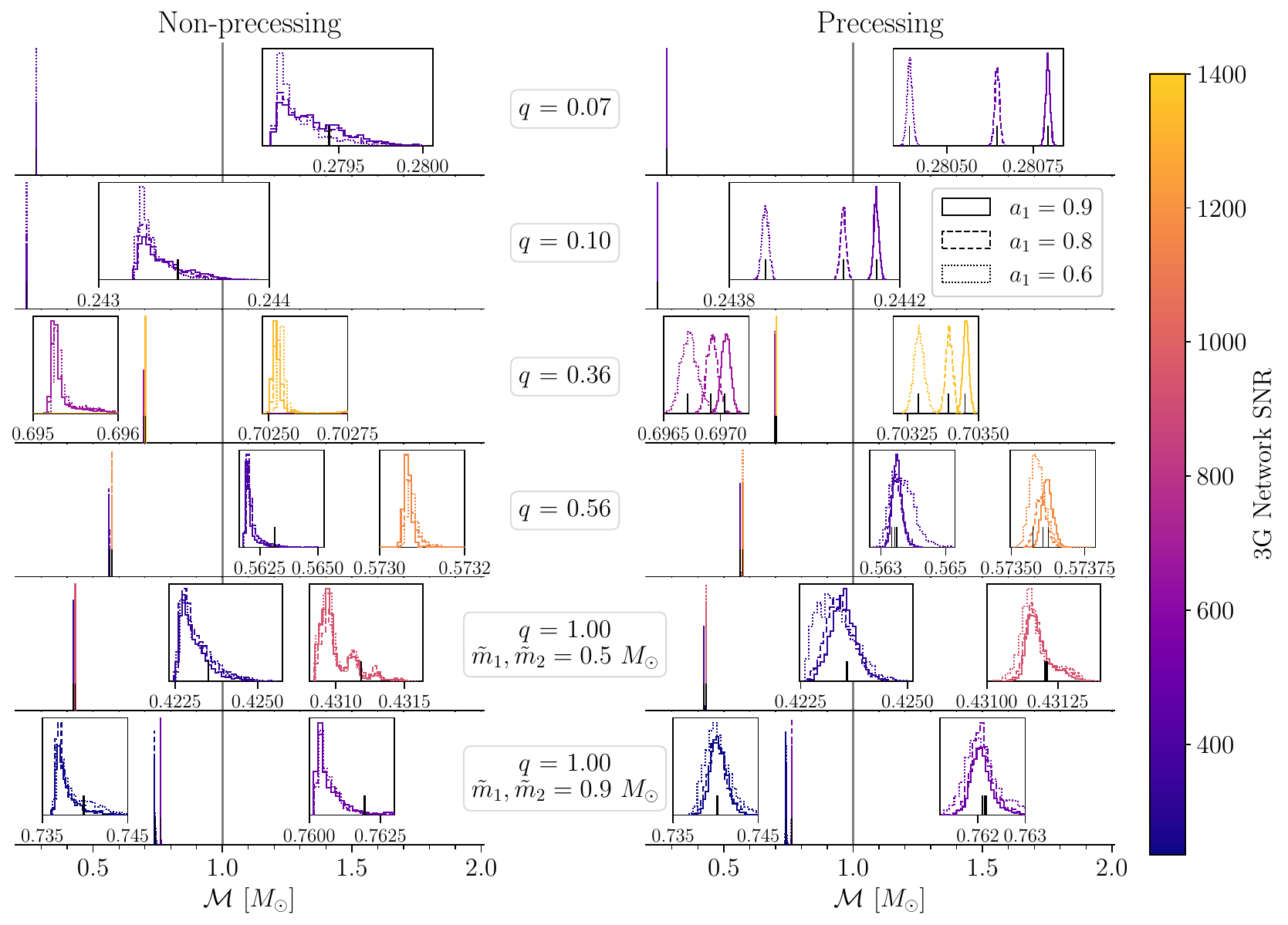}
    \caption{
    Marginal posterior distributions on the source-frame chirp mass $\mathcal{M}$ for the simulated signals injected into a network of Cosmic Explorer and the Einstein Telescope.
    Results for non-precessing ($\theta_1 = 0$) sources are shown on the left and those for precessing sources ($\theta_1 = \pi / 2$) on the right.
    The posteriors are colored by the network SNR of the signal.
    The linestyle of the posterior reflects the dimensionless spin magnitude of the more massive black hole, $a_1 = 0.6$ (dotted), 0.8 (dashed), or 0.9 (solid).
    Posteriors are organized by increasing mass ratio of the source, from top to bottom.
    Thin black lines rising from the $\mathcal{M}$-axis indicate the true value of $\mathcal{M}$, and a grey line is included at the fiducial mass scale of $1~\msun$.
    We note that these posteriors are not normalized so that they may be visualized together.}
    \label{fig:3g-mchirp}
\end{figure}
\begin{figure}[ht]
    \centering
    \includegraphics[width=\linewidth]{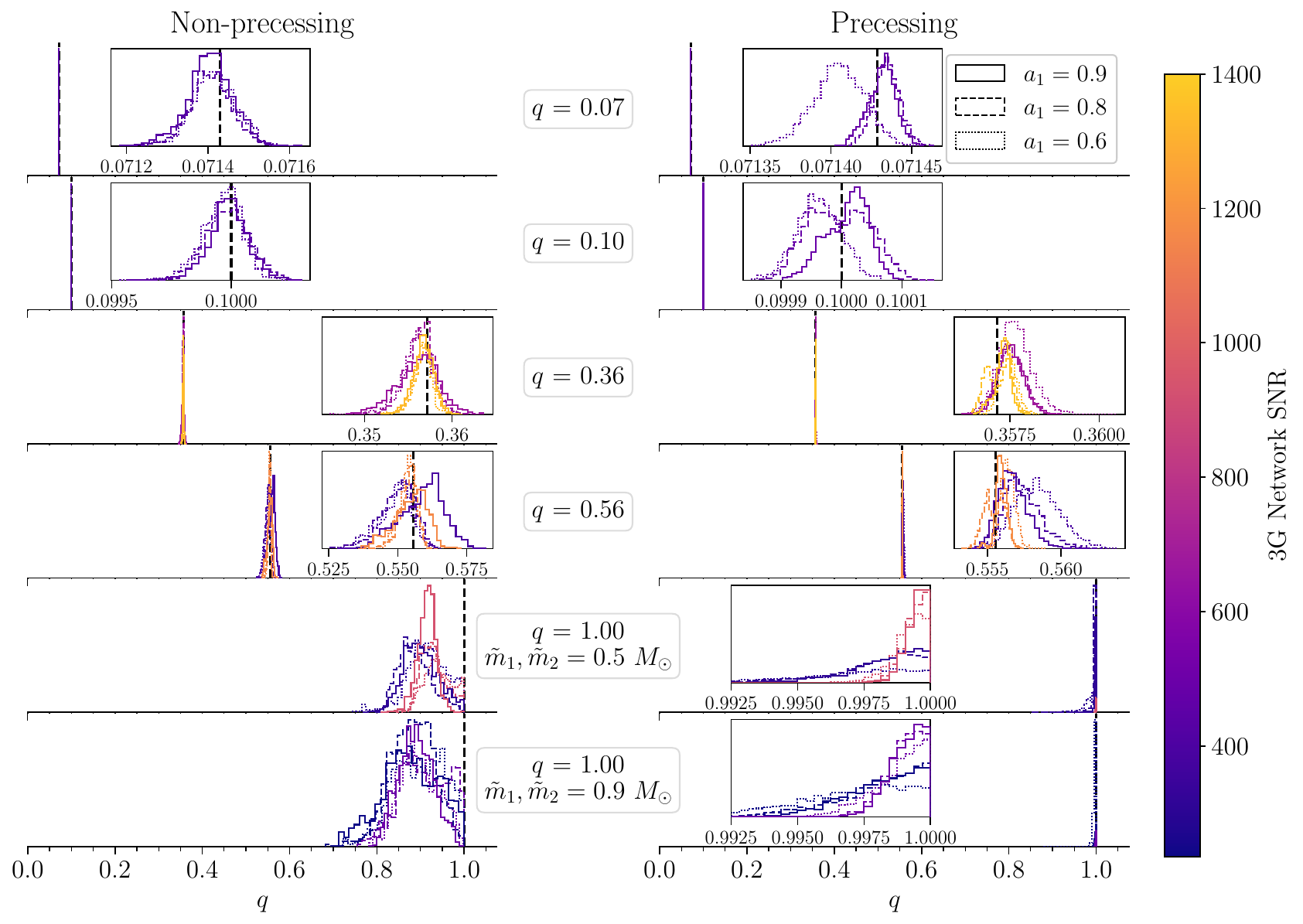}
    \caption{Marginal posterior distributions on the mass ratio $q$ for the simulated signals injected into a network of Cosmic Explorer and the Einstein Telescope.
    Results for non-precessing ($\theta_1 = 0$) sources are shown on the left and those for precessing sources ($\theta_1 = \pi / 2$) on the right.
    The posteriors are colored by the network SNR of the signal.
    The linestyle of the posterior reflects the dimensionless spin magnitude of the more massive black hole, $a_1 = 0.6$ (dotted), 0.8 (dashed), or 0.9 (solid).
    Posteriors are organized by increasing mass ratio of the source, from top to bottom.
    Dashed black lines indicate the true value of $q$.
    We note that these posteriors are not normalized so that they may be visualized together.}
    \label{fig:3g-massratio}
\end{figure}
\FloatBarrier
\begin{figure}[ht]
    \centering
    \includegraphics[width=0.5\linewidth]{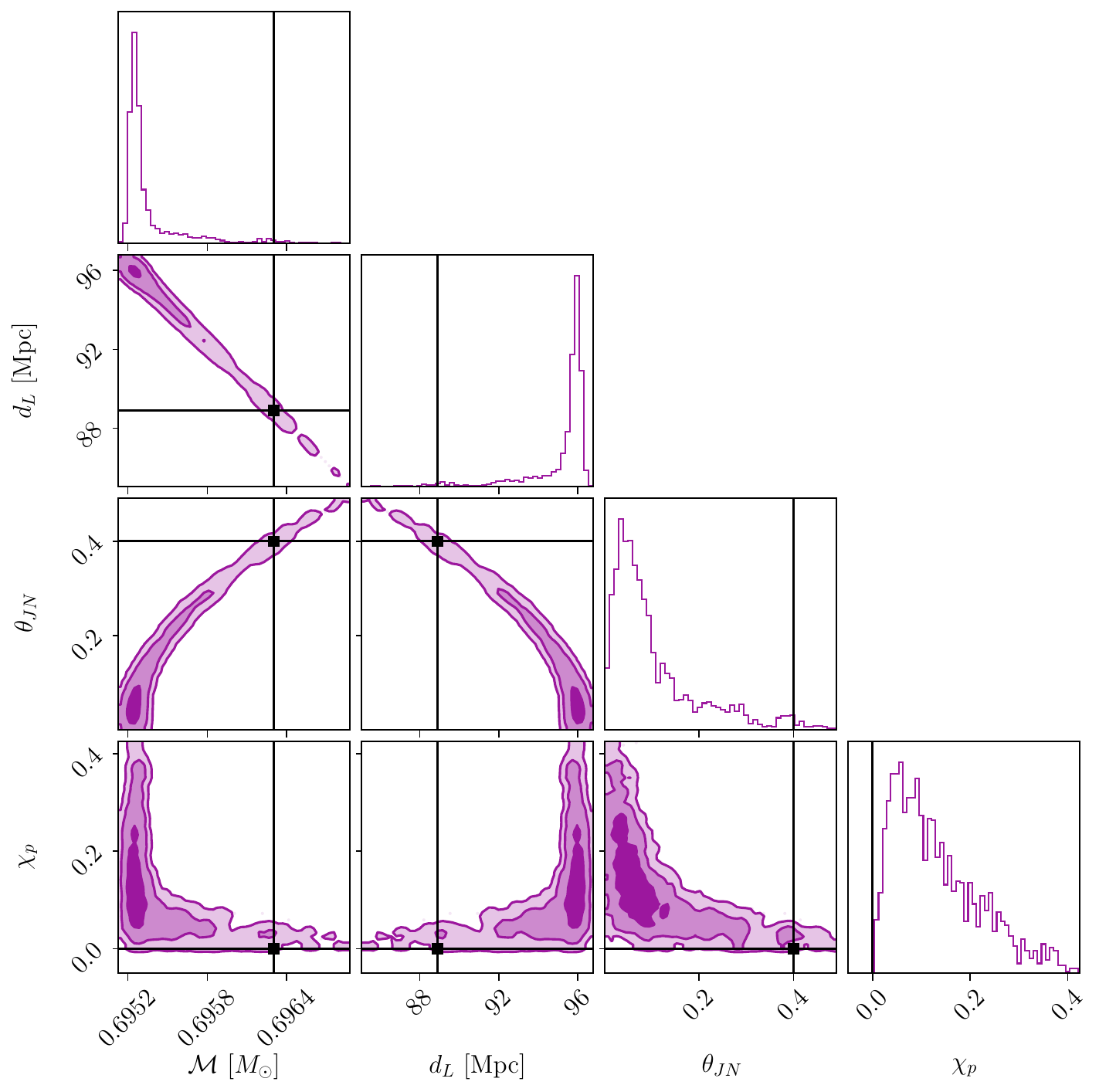}
    \caption{
    Marginal posteriors on the source-frame chirp mass $\mathcal{M}$, luminosity distance $d_L$, inclination angle relative to the line of sight \thetajn, and effective spin precession \chip for the non-precessing simulated signal with $q = 0.36$, $a_1 = 0.9$, and a network SNR of 665.5 in the XG network.
    True values are shown with black lines.
    We observe that the chirp mass is underestimated due to an overestimated distance, driven by correlations between $d_L$, \thetajn, and \chip.
    }
    \label{fig:3g-q0p36-a0p9-noprec-marginal}
\end{figure}
In Figures~\ref{fig:o4-mchirp} and \ref{fig:o4-massratio} we plot marginal posteriors on the source-frame chirp mass $\mathcal{M}$ and mass ratio $q$, respectively, for the signals simulated in the O4-design sensitivity network.
In Figures~\ref{fig:3g-mchirp} and \ref{fig:3g-massratio} we similarly plot marginal posteriors on $\mathcal{M}$ and $q$ for signals simulated in the XG network.
All of these figures are organized in the same manner as the plots of the marginal posteriors on the source-frame component masses in Section~\ref{sec:masses}.
We note that in Figure~\ref{fig:3g-mchirp} the marginal posteriors on $\mathcal{M}$ for the louder set of non-precessing $q = 0.56$ signals plus all non-precessing, $q = 0.36$ signals do not recover the true values of the chirp mass, peaking away from these value by $\lesssim 10^{-2}~\msun$.
We emphasize that this is an artifact of marginalizing a 15-dimensional posterior distribution, and not a meaningful bias, as the value of the log-likelihood at the true parameters versus the parameters with highest posterior probability differs by roughly unity.
The apparent offset in the source-frame masses is due to the fact that the luminosity distance posterior (needed to convert detector-frame mass to source-frame mass) only includes the true value at the edge of its tail, see Figure~\ref{fig:3g-q0p36-a0p9-noprec-marginal}.
In turn, the exact shape of the distance posterior is driven by the interplay between the correlation of some of the binary parameters (specifically: distance, inclination angle and \chip, see Appendix~\ref{app:spinprec-thetajn-correlation} below) and their non-trivial priors.
As shown in Figure~\ref{fig:chip-posteriors}, for a system with small mass ratio, the \chip prior is rather peaked away from zero, but due to correlations, that pushes \thetajn to smaller values, which in turn moves the distance to higher values.
\FloatBarrier

\pagebreak
\clearpage

\section{Effective Spin Constraints on Spin Magnitude} \label{app:a2-constraints-chieff}
\subsection{Spin-aligned Systems}
If we make assumptions about the spin geometry of the binary black hole system, we can constrain the spin of the lighter object $a_2$ under the most conservative assumptions about the spin of the heavier object, $a_1$.
First, we consider two black hole spins both aligned with the orbital angular momentum, or both in the opposite direction of the orbital angular momentum; here, we refer to both cases as ``spin-aligned".
In these cases,
\begin{equation}
    |\chieff| = \frac{a_1 + q a_2}{1 + q}.
\end{equation}
Rearranging in terms of $a_2$,
\begin{equation}
    a_2 = \frac{|\chieff| (1 + q) - a_1}{q},
\end{equation}
and notice that $a_1 \leq 1$, so,
\begin{equation} \label{eqn:a2cond}
    a_2 \geq \frac{|\chieff| (1 + q) - 1}{q}.
\end{equation}
Therefore, $a_2 > \ans$ when the right-hand side of Equation~\ref{eqn:a2cond} is greater than \ans.
This is fulfilled for
\begin{equation}
    |\chieff| > \frac{\ans q + 1}{1 + q}
\end{equation}
which we recognize as $|\chieff|$ for spin-aligned black holes with $a_1 = 1, a_2 = \ans$.
For $\ans = 0.4$ \citep{Hessels:2006ze}, this critical value of \chieff is maximized at $q = 1$ and we find that $\chi_{\rm eff, crit} = 0.7$.
So, for spin-aligned systems, $|\chieff| > 0.7$ implies that $a_2$ \textit{must} be larger than the maximal neutron star spin, assuming the spins are aligned.

\subsection{Spin anti-aligned Systems}
We can construct a similar constraint in systems where one black hole spin is aligned with the orbital angular momentum and one points in the opposite direction.
Here, we take the more massive object to spin in the direction of the angular momentum and the less massive object to spin opposite it (the constraint we construct turns out to be the same if these are switched).
Then,
\begin{equation}
    \chieff = \frac{a_1 - q a_2}{1 + q},
\end{equation}
and so,
\begin{equation}
    a_2 = \frac{a_1 - \chieff(1 + q)}{q} \geq -\frac{\chieff (1 + q)}{q}
\end{equation}
as $a_1 \geq 0$.
Notice, for $a_1 = 0$ and with the spin of the lighter object pointing opposite the orbital angular momentum, $\chieff < 0$.
Thus,
\begin{equation}
    a_2 \geq \frac{|\chieff| (1 + q)}{q},
\end{equation}
and similar to our analysis for spin-aligned systems, we that $a_2 > \ans$ when
\begin{equation}
    |\chieff| \geq \frac{\ans q}{1 + q}
\end{equation}
which we recognize as the value of $|\chieff|$ for $a_1 = 0, a_2 = \ans$.
This critical value of $|\chieff|$ is maximized for $q = 1$, yielding $\chi_{\rm eff, crit} = 0.2$.
Again, above this threshold, $a_2$ \textit{must} be larger than the maximal neutron star spin, assuming the spins are anti-aligned.

\section{Features in Correlated Mass Ratio-Effective Spin Posteriors} \label{app:chieff-q-bias}
\begin{figure}
    \centering
    \begin{tabular}{cc}
        \includegraphics[width=0.49\linewidth]{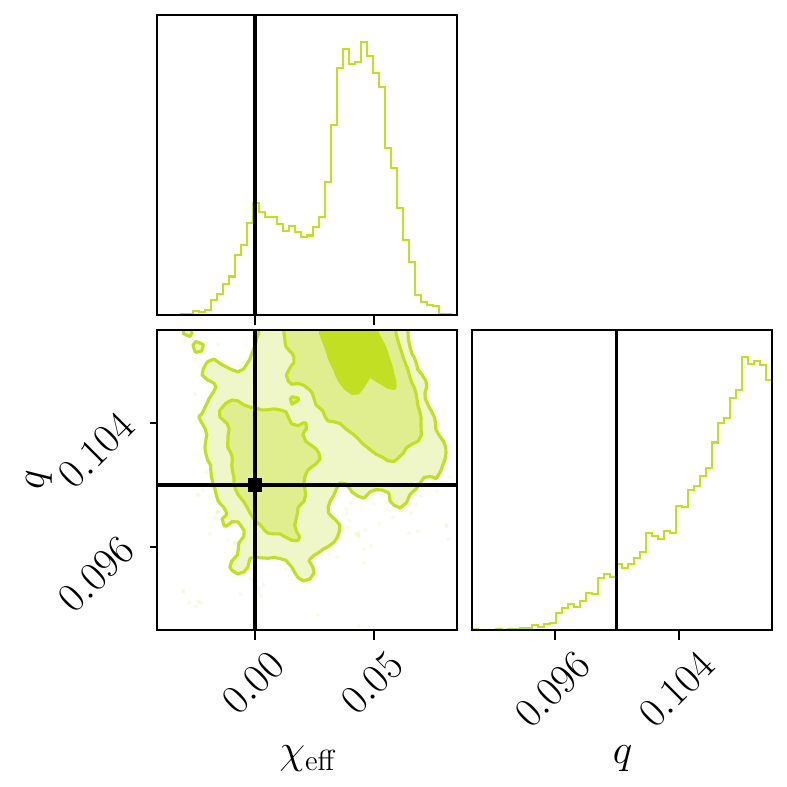} & \includegraphics[width=0.49\linewidth]{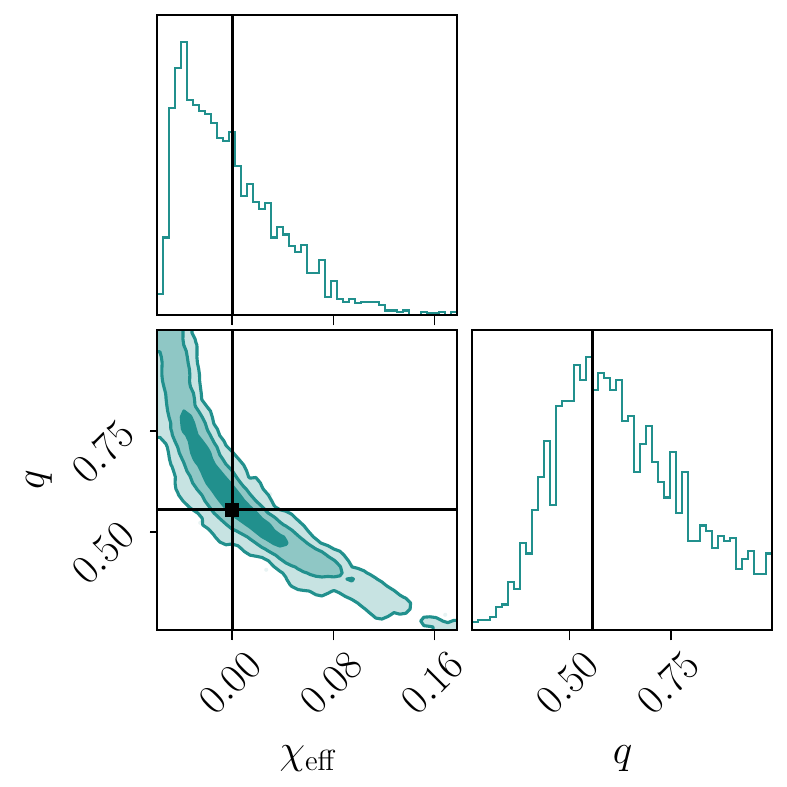} \\
    \end{tabular}

    \caption{
    Marginal posterior distributions on mass ratio $q$ and effective spin \chieff for two precessing signals (with $a_1 = 0.6$) in the O4 detector network that do not peak at the true value of \chieff, driven by the $q-\chieff$ correlation.
    True values are shown with black lines, and the distributions are colored by the mass ratio with $q = 0.10$ (left, lime) and $q = 0.56$ (right, green-blue), matching the colors of Figure~\ref{fig:chieff-posteriors}.
    The $q = 0.10$ run has a network SNR of 14.3, and the $q = 0.56$ run has an SNR of 12.5. 
    }
    \label{fig:chieff-q-posteriors}
\end{figure}

In Figure~\ref{fig:chieff-q-posteriors}, we show marginal posteriors for two precessing simulated signals with $a_1 = 0.6$ in the O4 detector network; the runs shown are at the lower SNRs listed in Table~\ref{tab:masses}.
The source with $q = 0.10$, which has a network SNR of 14.3 (left panel, lime color) shows bimodality in the posterior distribution for \chieff driven by posterior support for $q$ away from the true value, shown in black.
As $q$ and \chieff are correlated, a small change in $q$ drives samples to a new iso-likelihood ridge in the space of $q-\chieff$, introducing a new mode at larger \chieff.
In the 2D marginal $q-\chieff$ posterior for this source, the lower probability mode coincides with the true value.
It does not dominate the posterior because the likelihood is especially ``peaky" and, in a sense, ``underresolved".
As $q$ decreases, the length of the signal increases, and so even relatively minor deviations in the source parameters will result in a large mismatch between a proposal for the signal and the data; thus, the likelihood peaks around an especially narrow range of source parameters.
Using \code{dynesty}, the number of nested sampling live points $N_{\rm live}$ can be heuristically related to our resolution of the total prior volume.
If we have too few live points, it is less likely to initially place them within the narrow width of the maximum likelihood peak.
The analysis for the precessing $q = 0.10$, $a_1 = 0.6$ source in the O4 network employed $N_{\rm live} = 2000$.
Using fewer live points, the true value lay at the edge of the 1$\sigma$ level of the marginal posterior on $q$, and with increasing $N_{\rm live}$ we observe a (still subdominant) mode emerge centered on the true source parameters.
It is likely that increasing $N_{\rm live}$ further would allow us to fully resolve a unimodal posterior centered on the true $q$ and \chieff, although this becomes increasingly computationally expensive.

The source with $q = 0.56$, with a network SNR of 12.5 (right panel of Figure~\ref{fig:chieff-q-posteriors}, green-blue color) underestimates \chieff, below the true value of zero.
However, inspecting the joint posterior on $q-\chieff$ we see that the true value lies along a ridge of similar probability, making it (roughly) equally as likely for \chieff to be slightly negative, and with slightly larger $q$.
Without better measurement of the spin magnitudes and tilts, a larger range of \chieff is allowed.
Since the marginal posterior on $q$ shows additional support at larger values, we underestimate \chieff.

\section{Additional Marginal Posteriors on Spin Geometry} \label{app:extra-spins-posteriors}
\FloatBarrier

\begin{figure}[ht]
    \centering
    \includegraphics[width=\linewidth]{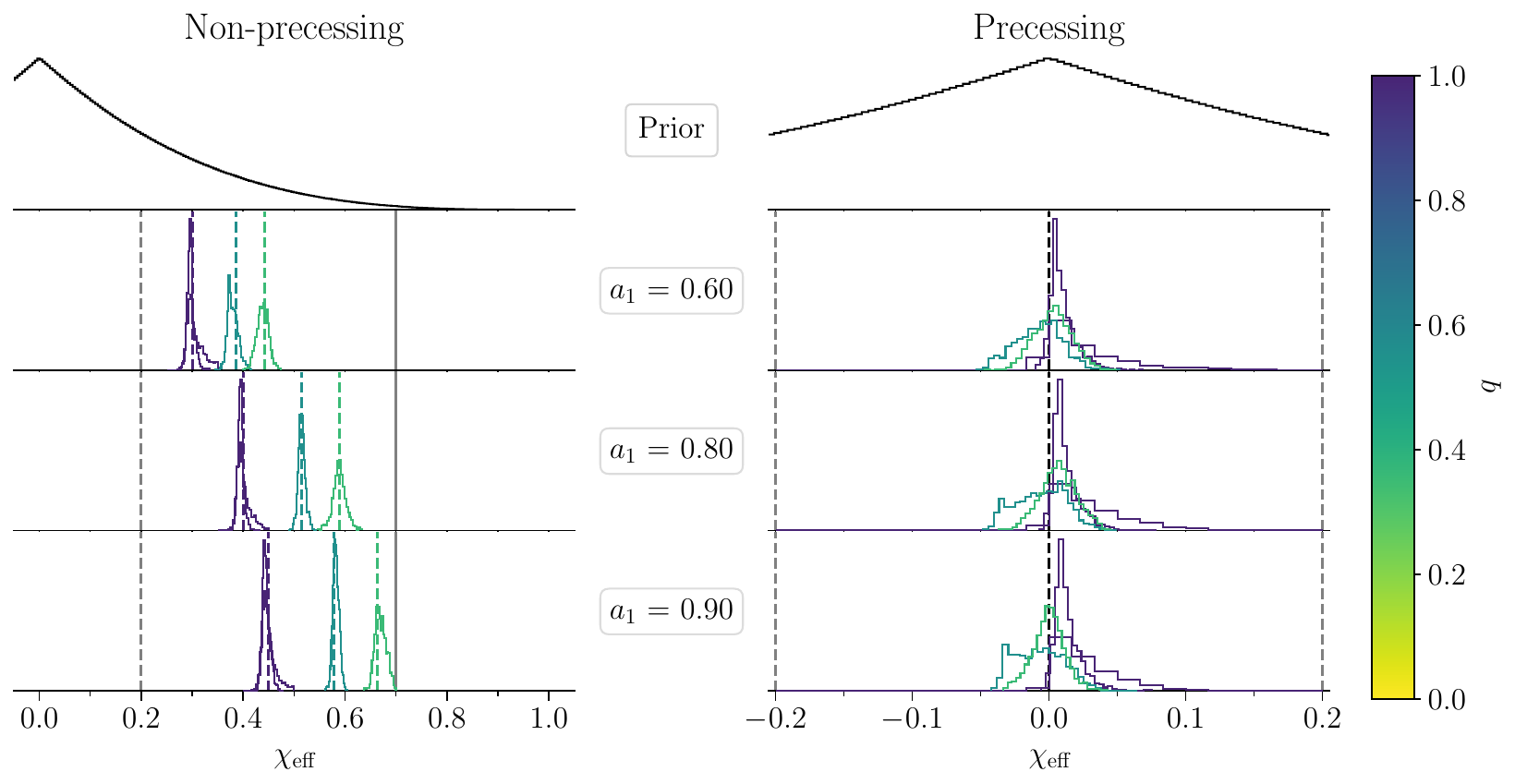}
    \caption{Marginal posterior distributions on \chieff for the louder simulated signals in the O4 detector network.
    This figure is organized in the same manner as Figure~\ref{fig:chieff-posteriors}, with the marginal posteriors colored by the true mass ratio $q$ and organized from top to bottom by the true value of $a_1$, with the prior on \chieff shown at the top.
    Non-precessing signals are on the left, with the true value of \chieff shown with a dashed line colored by the true $q$.
    Precessing signals are shown on the left where the truth is shown with a dashed black line.
    We also include the two critical values of \chieff above which $a_2$ is inconsistent with neutron star spin derived in Appendix~\ref{app:a2-constraints-chieff}, assuming that spins are aligned (solid grey) or anti-aligned (dashed grey).
    Note that there are no signals with $q \leq 0.1$ in this plot as those sources were only simulated with one SNR (see Table~\ref{tab:masses}).}
    \label{fig:highsnr-chieff-posteriors-o4}
\end{figure}
\begin{figure}[ht]
    \centering
    \includegraphics[width=\linewidth]{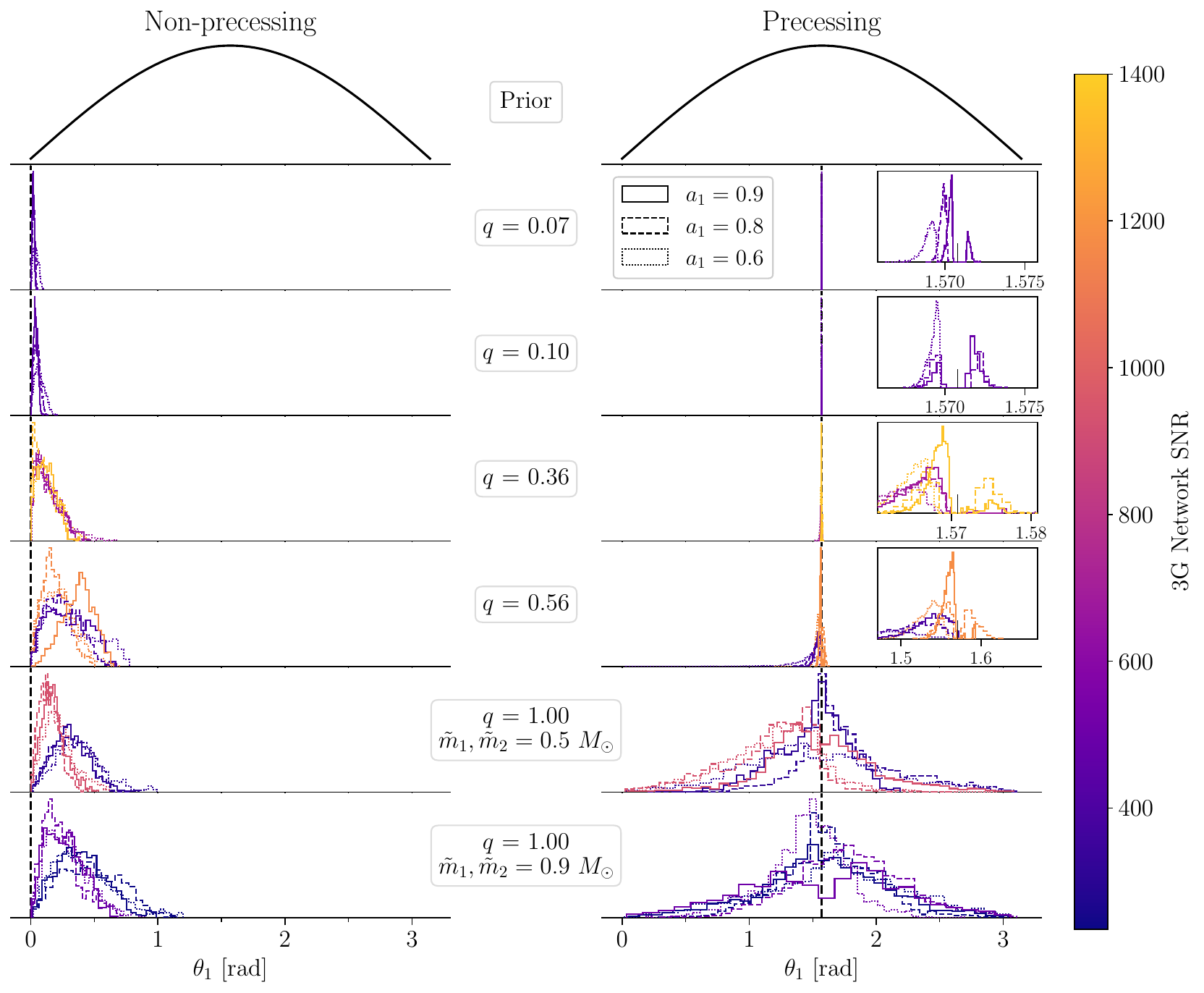}
    \caption{Marginal posterior distributions on the tilt angle between the spin vector of the heavier black hole and the orbital angular momentum, $\theta_1$, for the quietest signals for a given set of intrinsic parameters $(\tilde{m}_1, \tilde{m}_2, a_1, \theta_1)$, injected into a network of Cosmic Explorer and the Einstein Telescope.
    In the top row, we show the prior on $\pi(\theta_1)$.
    The truth is indicated with a dashed black line.
    We color the posterior distributions by the network SNR, and their linestyles reflect the true spin magnitude of the heavier black hole, $a_1$.
    We observe that $\theta_1$ is well-measured for non-precessing systems, and at low mass ratios.
    As noted in Section~\ref{sec:spins-3g}, the lack of posterior support at $\theta_1 = \pi / 2$ for precessing signals reflects the aligned-spin prior effect investigated in Appendix~\ref{app:tilt-prior}.}
    \label{fig:tilt1-posteriors-3g}
\end{figure}

\FloatBarrier

\section{Biased Tilt Posteriors at High Signal-to-Noise Ratios} \label{app:tilt-prior}
\begin{figure}[ht]
    \centering
    \includegraphics[width=0.5\linewidth]{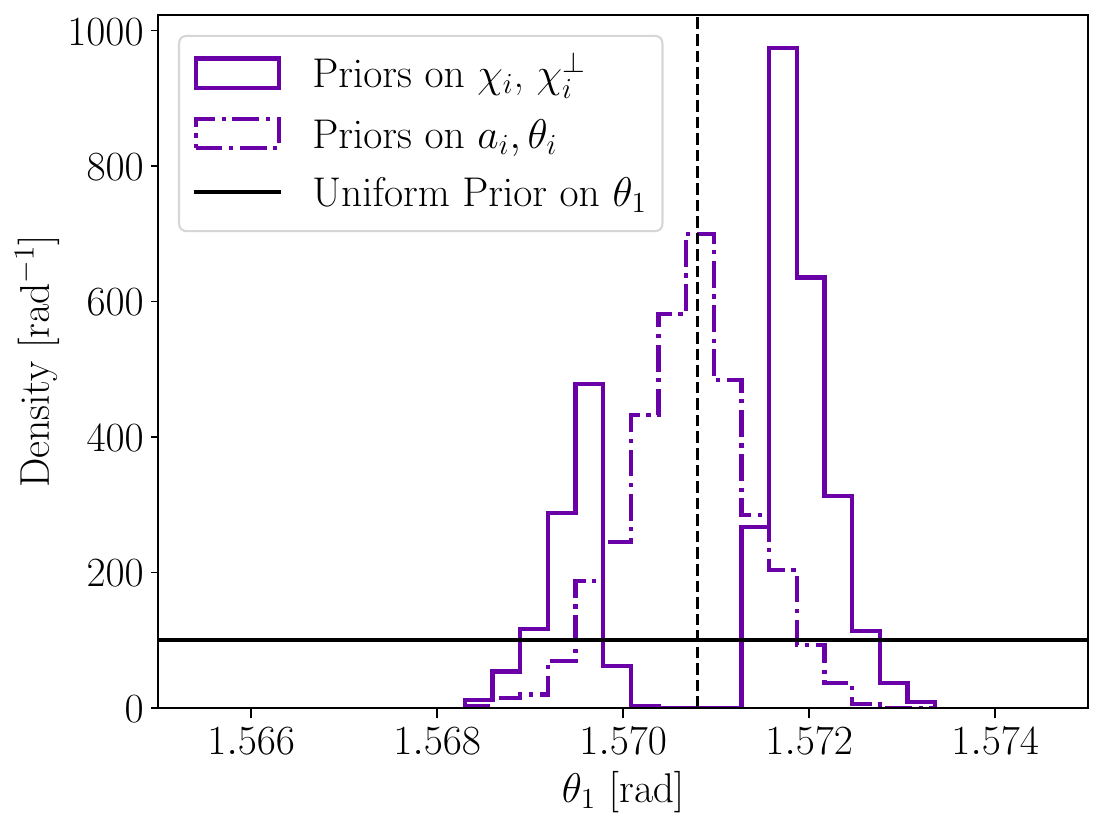}
    \caption{
    Marginal posterior on $\theta_1$ from two analyses of the simulated source with $q = 0.10$, $a_1 = 0.9$, and $\theta_1 = \pi / 2$ (dashed black line) injected into the XG network with an SNR of 490.3.
    Here, we repeat the posterior shown in Figure~\ref{fig:3g-tilt1-reduced-posteriors} (solid line) and compare it to the posterior generated when adopting uniform priors on the spin magnitudes $a_i$ and tilt angles $\theta_i$ (dash-dotted line).
    The uniform prior on $\theta_1$ is shown in the black solid line.
    We observe that sampling in the spin magnitudes and tilts directly recovers a posterior of similar width as the posterior recovered when sampling in the aligned ($\chi_i$) and in-plane spins ($\chi_i^\perp$), and it also recovers the true value of $\theta_1$.
    }
    \label{fig:3g-highsnr-tilt-prior-comparison}
\end{figure}

Assuming isotropic priors on the spin magnitudes and tilt angles, the prior on the aligned components of the spins $\pi(\chi_i)$ formally diverges to infinity at $\chi_i = 0$, i.e. $\theta_i = \pi / 2$ (see equation A7 of Ref.~\cite{Lange:2018pyp}).
Nested sampling in \code{dynesty} takes place in a unit hypercube, requiring the inverse of the cumulative distribution function (CDF) of the prior; however, the CDF of $\pi(\chi_i)$ is not analytically invertible.
Instead, \code{bilby} constructs a numerical approximant to the CDF which insufficiently resolves the divergence at $\chi_i = 0$ in $\pi(\chi_i)$ for very high SNR signals, resulting in little to no posterior support at $\theta_i = \pi / 2$ as observed in Figures~\ref{fig:3g-tilt1-reduced-posteriors} and \ref{fig:tilt1-posteriors-3g}.

To verify that this bias in $\theta_1$ is indeed a prior effect, we sample directly in the spin magnitudes and tilts to re-analyze the simulated signal with $q = 0.10$, $a_1 = 0.9$, and $\theta_1 = \pi / 2$ in the XG network of Cosmic Explorer and the Einstein Telescope, with a network SNR of 490.3.
In Figure~\ref{fig:3g-highsnr-tilt-prior-comparison} we compare the marginal posteriors on $\theta_1$ recovered by sampling in the aligned- and in-plane ($\chi_i^\perp$) spin components with isotropic priors (solid line) and in the spin magnitudes $a_i$ and tilt angles $\theta_i$ with uniform priors (dash-dotted line).
We note that, in the range of $\theta_1$ considered very near $\pi / 2$, this is nearly equivalent to an isotropic (sine) prior on $\theta_1$.
The posterior generated by sampling in $\chi_i$, $\chi_i^\perp$ is repeated from the inset panel in the second row, right column of Figure~\ref{fig:3g-tilt1-reduced-posteriors}.
Here, we see that the posterior generated by sampling in $a_i$, $\theta_i$ is of a similar width as the original result; sampling in $\chi_i, \chi_i^\perp$ we found a 90\% credible interval on $\theta_1$ of $3.3 \times 10^{-3}$ radians versus a credible interval of $2.2 \times 10^{-3}$ radians when sampling in the spin magnitude and tilt directly.
Importantly, we also successfully recover the true value of $\theta_1$ when using uniform priors on $a_i$, $\theta_i$, indicating that the bias in $\theta_1$ we observed in Figure~\ref{fig:3g-tilt1-reduced-posteriors} is a prior effect.

\section{Correlation Between Spin Precession and Binary Geometry} \label{app:spinprec-thetajn-correlation}
\begin{figure}
    \centering
    \includegraphics[width=0.5\linewidth]{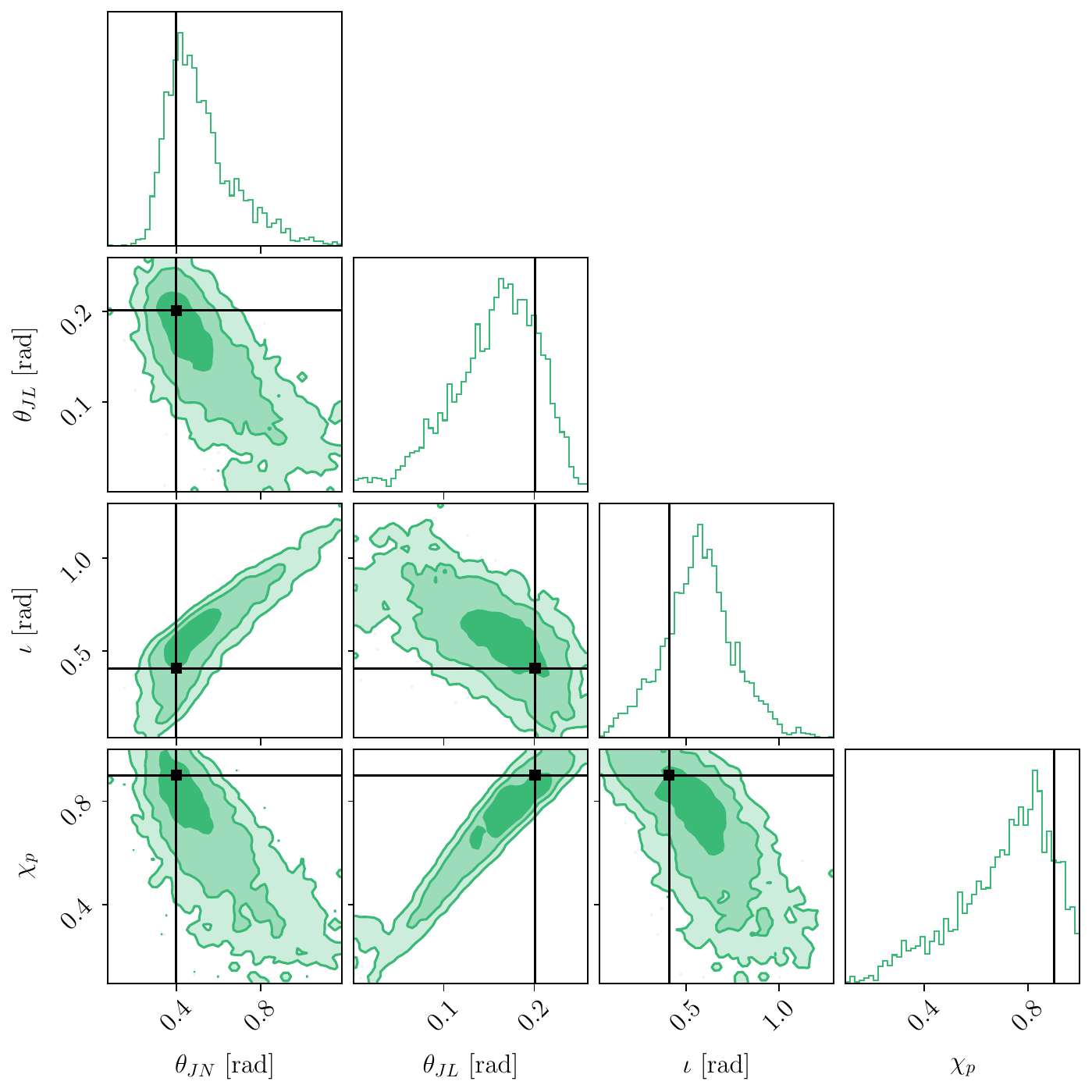}
    \caption{Marginal posteriors on the effective spin precession \chip and the zenith angles \thetajn (between the total angular momentum and line of sight), \thetajl (between the total and orbital angular momenta), and $\iota$ (between the orbital angular momentum and line of sight), from our analysis of the simulated $q = 0.36$, $a_1 = 0.9$ precessing signal in the O4-design sensitivity network.
    True values are shown in black lines.
    }
    \label{fig:spinprec-thetajn-correlation}
\end{figure}
At the three lowest mass ratios of the simulated precessing signals considered in this work, $q = 0.07, 0.10, 0.36$, we observed a correlation between the effective spin precession \chip and the zenith angles of the orbital angular momentum \vecL and total angular momentum \vecJ.
To our knowledge, this correlation has not been previously reported.
In Figure~\ref{fig:spinprec-thetajn-correlation} we show marginal posteriors on \chip, \thetajn (zenith angle between \vecJ and the line of sight $\hat{N}$), \thetajl (zenith angle between \vecJ and \vecL), and $\iota$ (zenith angle between \vecL and $\hat{N}$) for the precessing source with $q = 0.36$ and $a_1 = 0.9$ in the O4-design sensitivity network, which had a network SNR of 21.2.
We compute \thetajl (also referred to as the opening angle, $\beta$) at a reference frequency of 20 Hz using \code{pesummary.gw.conversions.spins.opening\_angle()}, which in turn implements methods from \code{LALSimulation}. 
Along the bottom row, we see linear correlations between \chip and each of these angles.

Heuristically, this correlation can be understood as follows: consider a gravitational-wave signal consistent with precession.
This necessarily implies a non-zero angle \thetajl between \vecL and \vecJ.
If we infer $\theta_1 \neq 0$, we can explain  
larger (smaller) values of this angle with larger (smaller) values of $a_1$ which pushes \vecJ into (away from) the orbital plane and away from (towards) \vecL.
Note that in all of the signals we simulated with $a_1 > 0, a_2 = 0$, we have $\chip = a_1 \sin \theta_1$ (c.f. Equation~\ref{eqn:chip}).\footnote{Alternatively, we could increase $\theta_1$ if we infer non-zero spin $a_1$ to achieve the same effect, although the change in \chip with respect to $\theta_1$ is much smaller near the true value of $\theta_1 = \pi / 2$ when $\chip \propto \sin \theta_1$.}
So, larger \chip necessitates larger \thetajl, and thus these quantities are positively correlated.
Alternatively, if we are uncertain of the degree to which we observe precession in a gravitational wave signal, we can fix all of the geometry of our system except for \thetajl and \thetajn, in particular fixing the component black hole spins and $\iota$.
As \vecL precesses about \vecJ, if we enlarge the cone of precession by increasing \thetajl then we must bring this cone closer to the line of sight by decreasing \thetajn to keep $\iota$ constant.
In this way, \thetajn and \thetajl are negatively linearly correlated.

Combined, we have a negative linear correlation between \chip and \thetajn.
Similar heuristic arguments can explain the observed correlations with $\iota$.
In detail, this effect is likely because precession for these systems is in the ``tropical region" noted by \citep{Apostolatos:1994mx}, where the plane of the orbit can wobble so violently that both the top and bottom are observed along the line of sight in the course of precession.
This injects additional information on the binary geometry of the system into the gravitational wave signal, which has previously been observed to reduce uncertainty in the measurement of \thetajn (see e.g. Figure~4 of \citep{Vitale:2014mka} or Figure~20 of \citep{Vitale:2016avz}).

\bibliographystyle{JHEP}
\bibliography{refs, refs-intro, refs-methods, refs-results, refs-software}

\end{document}